\documentclass[accepted]{melba}

\usepackage{mwe} 

\usepackage{amsmath,amssymb,amsfonts}
\usepackage{algorithmic}
\usepackage{graphicx}
\usepackage{textcomp}
\usepackage{multirow}
\usepackage{caption}
\usepackage{adjustbox}
\usepackage{array}
\usepackage{makecell}
\usepackage{booktabs}
\usepackage{nicematrix}
\usepackage{dsfont} 

\hypersetup{
    colorlinks=true,
    linkcolor=blue,
    citecolor=blue,
    urlcolor=blue
}

\newcommand{\FineRule}[1]{%
  \arrayrulecolor{gray!50}%
  \cmidrule(lr){#1}%
  \arrayrulecolor{black}%
}

\definecolor{modifcolor}{rgb}{0.0, 0.0, 0.0}
\newcommand{\modif}[1]{\textcolor{modifcolor}{#1}}

\melbaid{2026:032}  
\doi{10.59275/j.melba.2026-e93b}
\melbaauthors{Bernard}  
\email{olivier.bernard@insa-lyon.fr}
\volume{2026}
\firstpageno{658}  
\melbayear{2026}  
\datesubmitted{2025-12-122}  
\datepublished{2026-08-28}  

\ShortHeadings{The MYOSAIQ Challenge}{Bernard et al.}

\title{The MYOSAIQ Challenge: Myocardial Segmentation with Automated Infarct Quantification}

\author{
	\firstname Olivier \surname Bernard\aff{1,2}\orcid{0000-0003-0752-9946},
	\name William A. Romero R.\aff{1}\orcid{0000-0001-7312-6745},
    \name Cyprien Bouton\aff{1},
    \name Celia Goujat\aff{1},
    \name Hang Jung Ling\aff{1},
    \name Pierre-Marc Jodoin\aff{3},
    \name Fumin Guo\aff{4},
    \name Calder Sheagren\aff{5},
    \name Graham Wright\aff{5},
    \name Abdul Qayyum\aff{6},
    \name Moona Mazher\aff{7},
    \name Steven A. Niederer\aff{6},
    \name Hairui Wang\aff{7},
    \name Xiaomei Wu\aff{8},
    \name Franz Thaler\aff{9,10},
    \name Gernot Plank\aff{9},
    \name Martin Urschler\aff{11},
    \name Ricardo M. Rosales\aff{12}\orcid{0000-0002-8867-3802},
    \name Esther Pueyo\aff{12}\orcid{0000-0002-1960-407X},
    \name Nicolas Duchateau\aff{1,2}\orcid{0000-0001-8803-2004},
    \name Frederic Cervenansky\aff{1}\orcid{0000-0002-7817-1198},
    \name Patrick Clarysse \orcid{0000-0002-5495-7655},
    \name Loic Belle\aff{13}\orcid{0000-0003-0751-1949},
    \name Thomas Bochaton\aff{14,15}\orcid{0000-0002-5889-7506},
    \name Nathan Mewton\aff{14,15}\orcid{0000-0002-4526-8129},
    \name Magalie Viallon\aff{1,16}\orcid{0000-0001-9118-0438},
    \name Pierre Croisille\aff{1,16}\orcid{0000-0003-4019-3460}
}
\affiliations{
	\num 1 \addr  INSA-Lyon, Universite Claude Bernard Lyon 1, UJM-Saint Etienne, CNRS, Inserm, CREATIS UMR 5220, U1294, F-42023, Saint Etienne, France \\
	\num 2 \addr Institut Universitaire de France (IUF), France \\
	\num 3 \addr Department of Computer Science, University of Sherbrooke, Sherbrooke, QC, Canada\\
	\num 4 \addr Wuhan National Library of Optoelectronics, Huazhong University of Science and Technology, Wuhan, China\\
    \num 5 \addr Sunnybrook Research Institute, University of Toronto, Toronto, Canada\\
    \num 6 \addr National Heart and Lung Institute, Faculty of Medicine, Imperial College London, London, United Kingdom\addr \\ 
    \num 7 \addr Centre for Medical Image Computing, Department of Computer Science, University College London, London, United Kingdom \\
    \num 8 \addr Fudan University, Shanghai, 200433, China\\ 
    \num 9 \addr Gottfried Schatz Research Center: Medical Physics and Biophysics, Medical University of Graz, Graz, Austria\\ 
    \num 10 \addr Institute of Computer Graphics and Vision, Graz University of Technology, Graz, Austria\\ 
    \num 11 \addr Institute for Medical Informatics, Statistics and Documentation, Medical University of Graz, Graz, Austria\\ 
    \num 12 \addr Instituto de Investigaci\'on en Ingenier\'ia de Arag\'on, Universidad de Zaragoza; Instituto de Investigaci\'on Sanitaria de Arag\'on (IIS Arag\'on); CIBER-BBN, Zaragoza, Aragón, Spain\\ 
    \num 13 \addr Centre Hospitalier Annecy Genevois, France\\ 
    \num 14 \addr Hôpital Universitaire Cardiologique Louis Pradel, Bron, France\\
  \num 15 \addr  CarMeN INSERM U1060, Universite Claude Bernard Lyon 1,  Lyon, France\\
\num 16 \addr Hôpital Universitaire de Saint-Etienne, Saint-Etienne, France.
}

\abstract{
	Late gadolinium enhancement (LGE) cardiac magnetic resonance (MR) imaging is the modality of choice to assess myocardial infarction (MI) lesions. Nowadays MI volume quantification is not performed routinely in clinical practice. Numerous deep learning (DL) methods have been developed to automate the segmentation of the myocardium and infarct regions. However, most studies rely on relatively small datasets which typically undergo pre-processing steps to standardize images and focus on a specific phase of myocardial infarction following reperfusion therapy. These limitations have impeded the development of models that are generalizable across diverse conditions and thus suitable for routine clinical use. To advance research and establish benchmarks in generalizable learning for myocardial infarct quantification, this paper presents findings from the Myocardial Segmentation with Automated Infarct Quantification (MYOSAIQ) challenge. The dataset set up for the challenge combines 439 CMR volumes from two multicenter clinical trials, with representative data acquired in acute and chronic phases after acute MI. Data were acquired in 16 centers using MRI scanners from three different vendors. Six teams participated until the end of the challenge, employing various baseline models, data augmentation techniques, and confidence strategies. To enhance the significance of this study, we compare the challengers' results with those of fine-tuned foundation models. Our results indicate that well-designed UNet-based techniques outperform fully automatic foundation models for LGE MR segmentation. While the best methods achieve high-quality and stable delineations of the left ventricle and myocardium under various conditions, they remain improvable in accurately segmenting infarct regions.}


\begin{document}

\twocolumn[\maketitle]


\section{Introduction}
	\enluminure{L}{ate} Gadolinium Enhanced Cardiac Magnetic Resonance Imaging is a critical tool in clinical practice for evaluating myocardial damage following myocardial infarction (MI). This imaging technique is the only one enabling the quantification of myocardial tissue with abnormal gadolinium contrast washout, thereby highlighting areas of damaged myocardium. Accurately identifying these hyper-enhanced regions allows for precise determination of lesion size post-MI, a key factor for diagnosis, prognosis, and guiding therapeutic decisions
    \citep{ibanez_2019}. Moreover, the evolution of the size and features of the damaged tissue during the various inflammatory phases at the acute stage (within 8 days post-reperfusion) to the final infarct scar at the chronic stage (at least 1 month after percutaneous coronary intervention and tissue reperfusion) are key informations for patient recovery and follow-up characterization \citep{yellon_1983,ibanez_2019}. Thus, it is necessary to reliably estimate the lesion and infarct size at various stages over time. Finally, during the acute phase, blood flow may not be adequately restored into some regions of the myocardium despite successful reperfusion, resulting in post-reperfusion no-reflow lesions also called microvascular obstruction (MVO). Measuring the size of these no-reflow lesions is also crucial for understanding the progression of myocardial injury over time \citep{ibanez_2019}, but it also poses new challenges for segmentation \citep{Bulluck:OH:2016}.

    \modif{While quantification is clinically valuable and necessary, it is rarely performed in routine practice. This is mainly due to the difficulty of accurately segmenting MVO and infarct regions on LGE images. Indeed, contrast kinetics and imaging physics lead to blurred, time-dependent boundaries that are highly susceptible to artifacts, particularly at the infarct core–border zone interface. Moreover, the small size of these structures relative to the current image resolution further amplifies relative segmentation errors. In practice, the annotation process typically requires substantial manual input from clinicians to ensure accurate and consistent delineation of myocardial and lesion boundaries across all image slices.} As a result, automating this tedious and time-consuming task has become an active area of research, driven by recent advances in magnetic resonance (MR) image segmentation  \citep{chen_2020}. Indeed, the advent of the deep learning paradigm has motivated the development of several neural network-based techniques to improve late gadolinium enhancement (LGE) cardiac MR segmentation \citep{lalande_2022}. However, most of these techniques have been trained and evaluated on relatively small datasets ($\leq$150 patients), with cardiac imaging samples collected from a single clinical center. These studies typically rely on a unique 2D multi-slice, multi-breathhold LGE acquisition strategy, combined with pre-processing steps to standardize images, correct for motion artifacts between slices, and focus on a specific phase of myocardial infarction. While these studies serve as proof-of-concept and demonstrate the potential and feasibility of extracting meaningful MI features from LGE images under constrained conditions, none has yet been successfully translated into routine clinical practice or scaled to larger, more realistic datasets. In particular, they have not been validated on multicenter, multi-vendor cohorts comprising both acute and chronic cases, acquired using a broader range of LGE techniques and imaging protocols. It is therefore crucial to establish frameworks (i.e. databases and evaluation platform) that enable the study and refinement of the generalization capacity of the best segmentation methods.

    Open-access datasets strongly drive research progress on specific medical challenges. Earlier challenges (DE-MRI and MyoPS) on LGE segmentation have been conducted in small datasets ($\leq$45 patients) with images acquired under the same conditions \citep{Karim2016-jh,li_2023}. The recent LGE CMR dataset from the Evaluation of Myocardial Infarction from Delayed-Enhancement Cardiac MRI (EMIDEC) \citep{lalande_2022} stands as the most complete challenge and has been widely used to build and test the most advanced implementations of deep neural networks for the segmentation of cardiac structures in LGE CMRI. The top performing techniques \citep{lalande_2022} rely on a cascaded two-stage framework based on convolutional neural networks (CNN), where the left ventricle and the myocardium are first segmented, followed by the delineation of the MI lesion and the MVO \citep{lecesne_2023}. The obtained results show high segmentation scores for the left ventricle and the myocardium with encouraging but improvable scores for MI lesions. However, the EMIDEC dataset collected in a single center contains only 100 patients with unstaged lesions, and using the same LGE imaging technique. A pre-processing procedure was also applied to compensate for the motion artifacts between non-contiguous 2D slices.

    In this context, the MYOcardial Segmentation with Automated Infarct Quantification (MYOSAIQ) challenge was proposed and organized as part of the Functional Imaging and Modeling of the Heart (FIMH) conference in 2023 (endorsed by the MICCAI society), and was extended to a second phase in 2024. This challenge was set up to evaluate and compare the performance of automatic DL methods for quantifying myocardial LGE lesions observed after ischemia and reperfusion, at different phases of the longitudinal evolution of the disease. MR data were acquired during the acute phase and at delayed or chronic phases. Acute phase  refers to 4-8 days post-MI and reperfusion therapy, where gadolinium contrast distributes over the reperfusion lesion, highlighting an area mixing oedema, apoptotic or already necrotic tissue while also revealing MVO lesions as hypo-enhanced areas. Delayed or chronic phases refer to 1 month and 12 months post-MI and reperfusion, where the gadolinium contrast aggregates in the resulting final necrosis. Each data was annotated from a consistent contouring standard operating procedure described in section \ref{sec:data-annotation-procedure}. 
    
    In this context, the main contributions of this collaborative paper are:

    \begin{enumerate}
        \item We release a new fully annotated dataset comprising 439 LGE MR images specifically designed for myocardial infarct quantification. The value of this dataset lies in its heterogeneity including the main LGE variants imaging techniques, acquired in 16 centers with 3 MR vendors, without any pre-processing steps, and obtained either in the acute phase or at later chronic time points.
        \item We present the results from six distinct teams that participated in the MYOSAIQ challenge, each employing various baseline models, data augmentation techniques, and confidence strategies.
        \item We compare the challengers' results with those of foundation models for LGE MR images based on the well-known SAM architecture \citep{kirillov_2023}. These foundation models were specifically designed to address the generalization problem, yet their performance has not been comprehensively compared to state-of-the-art methods for myocardial infarction quantification. While these models are semi-automatic and require initialization—which may introduce bias into the analysis—we found it worthwhile to include this comparison to better contextualize the potential of such approaches.
        \item We conduct a comprehensive analysis considering multiple aspects, including overall segmentation quality, the impact of time following reperfusion therapy, the comparison between CNN and foundation models, and the effectiveness of confidence strategies.
    \end{enumerate}

\section{Challenge framework}

\subsection{The MYOSAIQ data collection}

The complete dataset includes 439 LGE examinations from two distinct multicenter clinical cohorts, designed to quantify MI lesions across different phases of disease progression. \modif{These phases encompass assessments conducted 4–8 days [D8] post-MI at the acute phase (MIMI cohort) and at 1 month [M1] and 12 months [M12] post-MI and reperfusion (HIBISCUS cohort).} Consequently, the dataset is divided into three subgroups corresponding to these phases:

\begin{itemize}
    \item 
    D8: 123 patients (98 for training, 25 for testing\modif{, using a random split}) with LGE images depicting acute myocardial infarction within 8 days post-MI (MIMI cohort). These patients underwent treatment via two different procedures: i) Percutaneous Coronary Intervention (PCI) with immediate stenting, or ii) the Minimalist Immediate Mechanical Intervention approach \cite{belle2016comparison}.
    \item 
    M1: 187 (155 for training, 32 for testing\modif{, using a random split}) patients with LGE images acquired at 1 month post-PCI and reperfusion (HIBISCUS-STEMI cohort).
    \item 
    M12: 129 (105 for training, 24 for testing\modif{, using a random split}) patients with LGE images captured at 12 months post-PCI and reperfusion (HIBISCUS-STEMI cohort).
\end{itemize}

\subsubsection{The MIMI cohort}

The MIMI cohort (ClinicalTrials.gov Identifier: NCT01360242) is derived from a multicenter randomized trial involving over 16 centers, designed to compare immediate stenting with 24–48-hour delayed stenting in patients undergoing percutaneous coronary intervention \citep{belle2016comparison}. The MR protocol included T1, T2, T2*, Cine, and 3D/2D LGE scans. The MIMI cohort consists of patients admitted to French hospitals with ST-segment elevation myocardial infarction (STEMI) within less than 12 hours of symptom onset. These patients were randomized into two groups: immediate stenting ($n=65$) or delayed stenting ($n=58$) following the restoration of Thrombolysis In Myocardial Infarction (TIMI) grade 3 flow via thrombus aspiration. Patients in the delayed stenting group underwent a second coronary angiography for stent implantation, with a median delay of 36 hours (interquartile range: 29–46 hours) after randomization. The primary endpoint evaluated was MVO, expressed as a percentage of left ventricular myocardial mass, assessed using cardiac magnetic resonance imaging performed 5 days after the initial procedure (interquartile range: 4-6 days). Although a trend towards lower microvascular obstruction was observed in the immediate stenting group compared to the delayed stenting group ($1.88\%$ versus 3.96\%; $P=0.051$), this trend became significant after adjusting for the area at risk ($P=0.049$) \citep{belle2016comparison}. 

\subsubsection{The HIBISCUS-STEMI cohort}
The HIBISCUS-STEMI cohort ClinicalTrials.gov
Identifier: NCT03070496) consists of consecutive patients admitted to the University Hospital of Lyon (HCL), a tertiary referral center, with suspected acute STEMI between 2016 and 2019. A total of 300 patients were included, all of them provided written informed consent. STEMI was defined according to the European Society of Cardiology guidelines, based on the presence of clinical symptoms (chest pain) associated with ST elevation $>2$ mm in two contiguous leads on a standard 12-lead electrocardiogram, along with a significant elevation in troponin-I levels. All patients underwent coronary angiography upon admission, followed by reperfusion via primary percutaneous coronary intervention. A complete myocardial enzyme release assessment was performed for all patients, who also underwent LGE cardiac MR imaging 1 month and 12 months after the acute myocardial infarction. All individual clinical, treatment, and outcome data were prospectively stored. Adverse clinical events were recorded during follow-up visits scheduled at 1 month, 1 year, and 2 years after the hospitalization. The MIMI and HIBISCUS-STEMI cohort were both approved by institutions's Review Board and Ethics Committee.

\subsubsection{LGE imaging protocols}

For LGE imaging in cardiac MR, the imaging protocols typically rely on inversion-recovery (IR) prepared T1-weighted sequences, usually performed 10 minutes after the intravenous administration of gadolinium chelate. When the sequence settings are optimized, a five-fold difference in intensity can be achieved between viable and nonviable myocardium \citep{amado2004accurate}. The recommended and most commonly used sequences were established as follows: (i) two-dimensional (2D) IR-prepared segmented gradient-echo (GRE) with magnitude and phase-sensitive inversion recovery (PSIR) reconstruction, and (ii) the three-dimensional (3D) IR-prepared segmented GRE with magnitude reconstruction \citep{viallon2011head}. The 2D-PSIR sequence is acquired over multiple breath-holds, which may be inconsistent in some patients (e.g., due to varying lung volume), whereas the 3D-IR-GRE sequence is acquired over a single, but longer, breathold. When feasible in patients, the latter also results in a higher contrast-to-noise ratio (CNR) between the gadolinium enhanced lesion and the normal myocardium, being the most performant sequence for LGE quantification \citep{viallon2011}. The MYOSAIQ dataset included $13\%$ and $87\%$ of 2D-PSIR and 3D-IR-GRE sequences, respectively. The sequence type information is not disclosed to participants to encourage the generalization of the deployed methods. All LGE cardiac MR images were acquired in DICOM format using 1.5 Tesla whole-body clinical multivendors MRI scanners. An in-house software was then used to transform the images into the NIfTI format (3D images), representing the final files delivered to the challenge participants.

\subsection{Data annotation procedure}
\label{sec:data-annotation-procedure}

\begin{figure}[t]
\centerline{\includegraphics[width=\columnwidth]{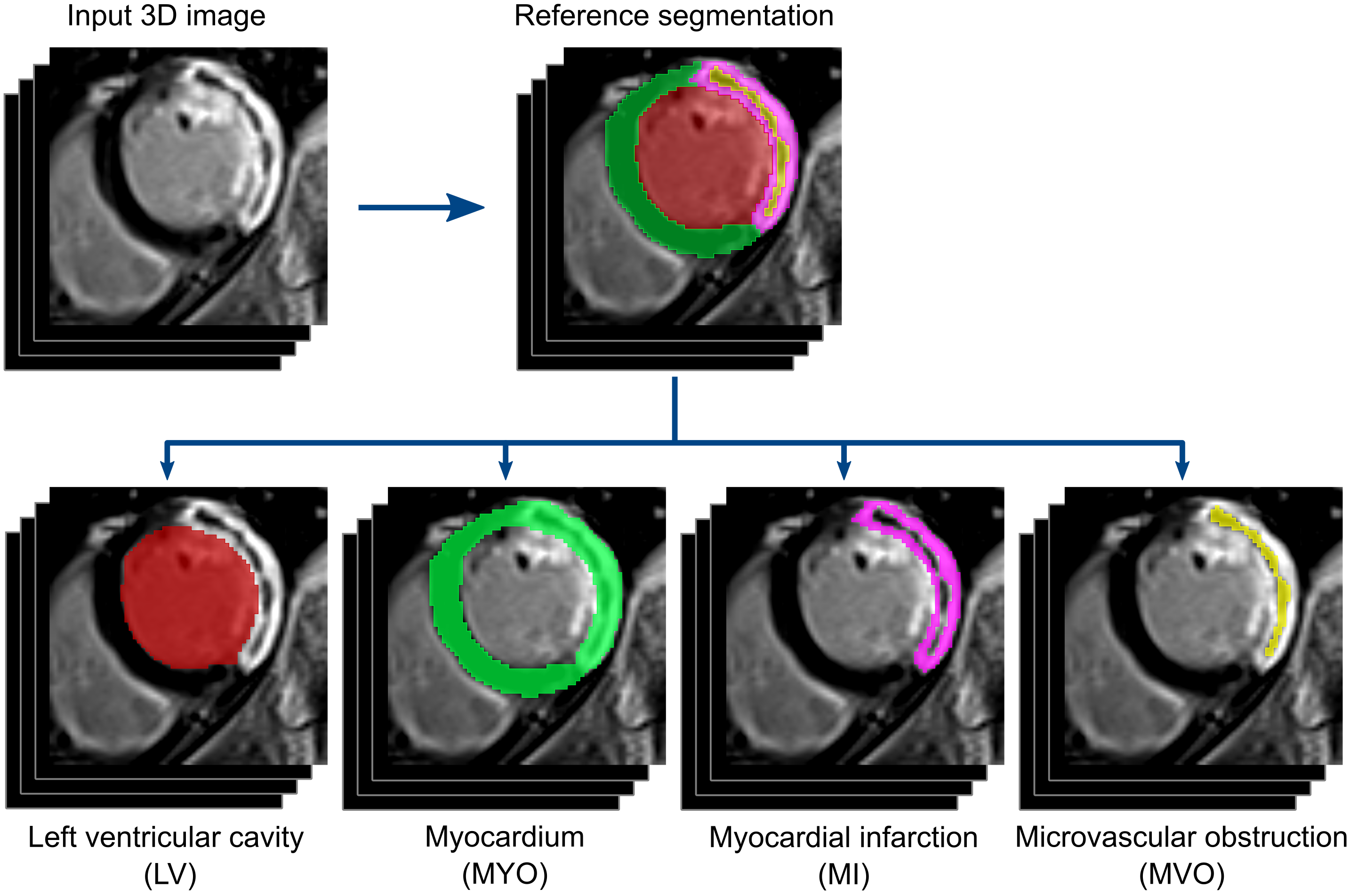}}
\caption{Reference segmentation and regions of interest: left ventricular cavity (red), myocardium (green), myocardial infarction (magenta) and microvascular obstruction (yellow).}
\label{fig:annotation}
\end{figure}

Every LGE cardiac MR image was manually annotated using the 3D Slicer software. An expert with over 20 years of experience in cardiac radiology and MRI performed the annotations for the entire dataset. To enable the evaluation of inter-observer variability, a second expert with 10 years of experience independently annotated the test set. The semi-automatic Full Width at Half Maximum (FWHM) approach currently recommended in the literature guided the experts but results were always systematically supervised and corrected if needed during the expert annotation process. Four main regions were considered: the left ventricle (LV) cavity, the left ventricle myocardium (MYO), the myocardial infarction (MI) and the microvascular obstruction (MVO) as presented in Figure~\ref{fig:annotation}. As such, the MYO corresponds to the entire myocardium, including both the healthy tissue and the regions affected by both MI and MVO when present. Particular attention was given to the delineation of each structure, especially at the base and apex, where the contours are prone to variations among experts \citep{bernard_2018}. To generate consistent annotations for the research community, we chose to apply the standard operating
procedures that were already used by the EMIDEC challenge, which rely on the following constraints:
\begin{itemize}
    \item The LV cavity must be completely covered, including the papillary muscles,
    \item MI refers to tissue death (infarction) of the heart muscle, and must be located inside the myocardium,
    \item MVO is characterized as hypo-enhanced regions, appearing as black areas surrounded by a bright rim within the myocardium. During the contouring of this structure, black signals caused by noise or artifacts should be excluded.
\end{itemize}

The reference-labeled segmentations were stored in the NIfTI format with the following labels: \mbox{background = 0}, \mbox{LV cavity = 1}, \mbox{healthy myocardium = 2}, \mbox{MI = 3}, and MVO = 4.

\subsection{Image-based clinically relevant descriptors}
\label{sec:clinical-properties}

\begin{figure}[t]
    \centering
    \includegraphics[width=1\linewidth]{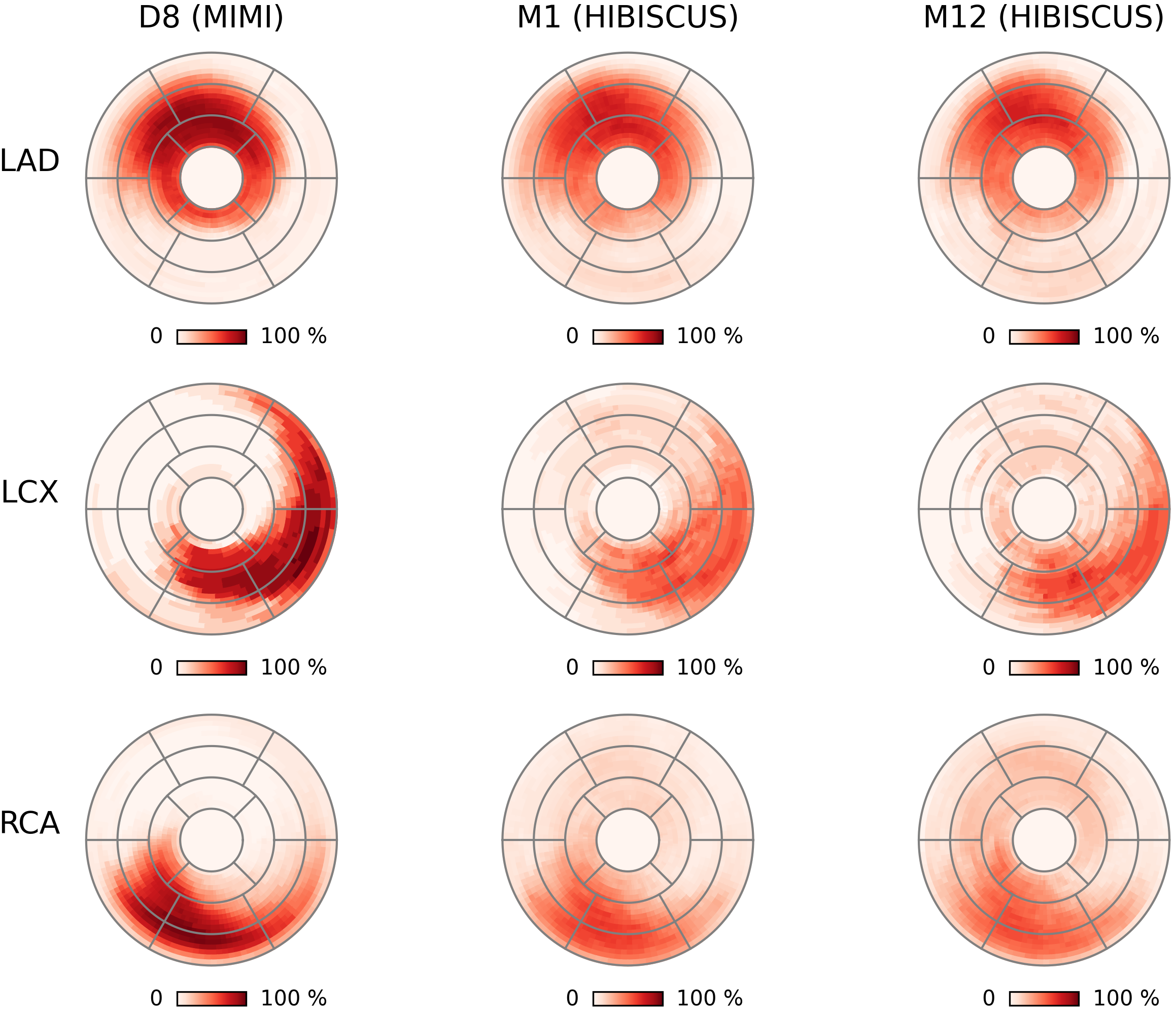}
    \caption{Infarct location of the LAD, LCX and RCA infarcts after 8 days, 1 month and 1 year post reperfusion (average across the subjects in each subgroup).}
    \label{infarct-location}
\end{figure}

To describe the properties of the MYOSAIQ dataset, Table~\ref{patient-table} presents the key descriptors of myocardial infarct lesions derived from expert annotations. These descriptors are categorized based on the affected myocardial territory—left anterior descending (LAD), left circumflex artery (LCX), and right coronary artery (RCA)—as well as the time points following reperfusion therapy (day 8 [D8], month 1 [M1], and month 12 [M12]). The extent of myocardial damage is expressed as a proportion of the total myocardium or the total endocardial surface area. In addition, Figure \ref{infarct-location} provides an overview of the mean infarct location for each affected myocardial territory and time point. For each subgroup, the corresponding Bull’s Eye display was computed using the method described in \citep{duchateau2023pixel}. 

\begin{table}[t]
\begin{center}
\setlength{\tabcolsep}{2pt} 
\resizebox{\columnwidth}{!}{%
\begin{tabular}{c c|c|c|c|c}
&& \textbf{Territory} & \textbf{D8} & \textbf{M1} & \textbf{M12} \\ \hline
\multicolumn{1}{l}{\textbf{Clinical metrics}} &  &  &  &  \\ 
\multicolumn{1}{c}{Transmurality}&{} & LAD &  $40 \pm 16$ & $27 \pm 15$ & $25 \pm 16$ \\ 
\multicolumn{1}{c}{\scriptsize(Extent of a lesion across)} & {(\%)} & LCX & $29 \pm 09$ & $17 \pm 09$ & $12 \pm 08$ \\ 
\multicolumn{1}{c}{\scriptsize the thickness of the muscle)}&{} & RCA & $24 \pm 10$ & $17 \pm 09$ & $12 \pm 08$ \\[5pt]
\multicolumn{1}{c}{}&{}& LAD & $34 \pm 11$ & $29 \pm 13$ & $24 \pm 14$ \\ 
\multicolumn{1}{c}{Endocardial Surface} & {(\%)}  & LCX & $33 \pm 08$ & $22 \pm 12$ & $19 \pm 11$ \\ 
\multicolumn{1}{c}{Length}&{} & RCA & $31 \pm 09$ & $18 \pm 09$ & $16 \pm 08$ \\[5pt]
\multicolumn{1}{c}{}&{} & LAD & $34 \pm 12$ & $18 \pm 11$ & $17 \pm 12$ \\ 
\multicolumn{1}{c}{Lesion Volume}&{(mL)} & LCX & $30 \pm 10$ & $13 \pm 08$ & $12 \pm 08$ \\ 
\multicolumn{1}{c}{}&{} & RCA & $23 \pm 14$ & $11 \pm 07$ & $11 \pm 06$ \\[5pt]  
\multicolumn{1}{c}{}&{} & LAD & $33 \pm 10$ & $19 \pm 10$ & $18 \pm 09$ \\ 
\multicolumn{1}{c}{Lesion Size}&{(\%)} & LCX & $25 \pm 07$ & $14 \pm 07$ & $11 \pm 06$ \\ 
\multicolumn{1}{c}{}&{} & RCA & $27 \pm 08$ & $14 \pm 09$ & $12 \pm 07$ \\ \\ \hline
\end{tabular}
}
\caption{Descriptors of the myocardial infarction lesion categorized by territory and time post reperfusion therapy. The endocardial surface length and the lesion size are provided in proportion of the whole endocardial length and the myocardial area, respectively.}
\label{patient-table}
\end{center}
\end{table}

\begin{figure}[htbp]
    \centering
    \includegraphics[width=1\linewidth]{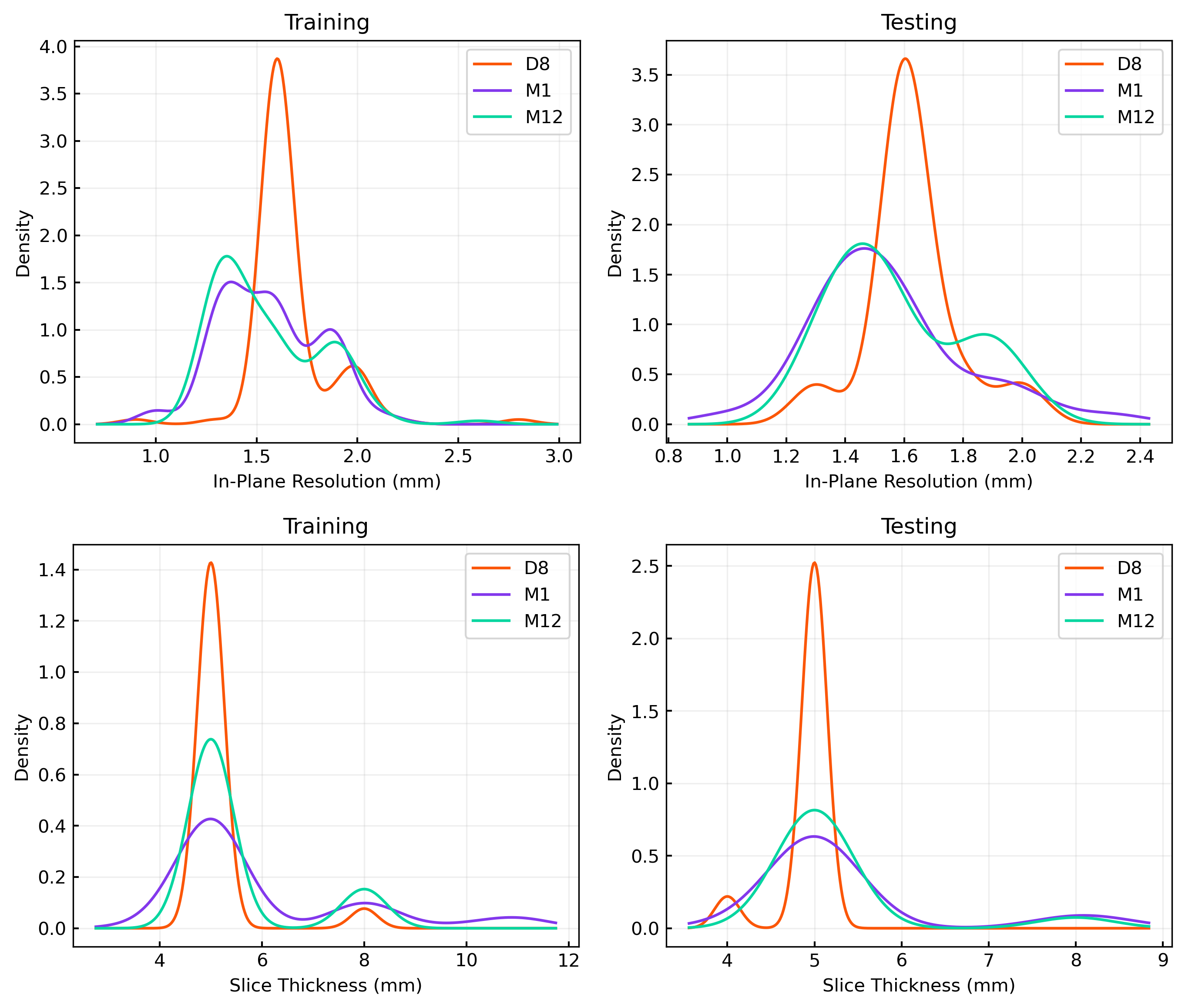}
    \caption{\modif{Distribution of the in-plane resolution and slice thickness for the training and test datasets across the three experimental configurations (D8, M1, and M12).}}
    \label{fig:spatial-distributions}
\end{figure}

\modif{Finally, Figure \ref{fig:spatial-distributions} illustrates the distributions of the in-plane resolution and slice thickness for the three experimental configurations (D8, M1, and M12), separately for the training and test datasets. As shown in the figure, the spatial resolution distributions remain highly comparable across all configurations. The in-plane resolution is consistently distributed between approximately 1.3 and 1.9 mm, while the slice thickness is centered around 5 mm in both the training and test datasets. These observations indicate that the spatial sampling characteristics are well matched between the training and test datasets, thereby limiting potential differences related to voxel spacing.}

\subsection{Evaluation platform}

A Codalab environment \citep{codalab_competitions_JMLR} was deployed as the evaluation platform for the segmentation challenge, offering a robust and transparent system for managing and assessing participants' submissions. The leaderboard allowed participants to monitor their progress and benchmark their results against others in real-time during both the training and testing phases. During the training phase \modif{(4 months)}, participants could evaluate their segmentations using the training dataset and corresponding reference masks, providing an opportunity for learning and experimenting with the platform. No restriction was imposed on the number of submissions during this phase, enabling extensive refinement of methods. In contrast, the testing phase \modif{(1 month)} limited participants to a maximum of five submissions on the testing dataset, with no access to the reference masks, ensuring a fair and competitive environment. The scoring code used on the online platform was released to all participants two months prior to the competition's launch. The source code, developed in Python, was published in a public repository\footnote{https://github.com/creatis-myriad/myosaiq-challenge} to ensure transparency and provide participants with sufficient time to familiarize themselves with the evaluation criteria and metrics. This platform is now accessible and will remain active and maintained as long as the data remains relevant for clinical research. A complementary set of metrics was employed to evaluate the quality of the automatically segmented masks in comparison to the ground truth, as detailed below:

\subsubsection{Geometrical metrics}

The standard Dice, the maximum Hausdorff surface distance (HD) and Average Symmetric Surface Distance (ASSD) were used to compute geometrical scores for the four labels, i.e. the LV cavity, the full myocardium, the MI and the MVO. All these metrics were computed from the 3D masks.

\subsubsection{Quality control metrics}
\label{sec:metrics}

We also implemented four indices to assess the quality of the estimation of the myocardial infarction lesion metrics, namely the correlation (CC), the mean absolute distance (MAE), the limits of agreement (LOA) and the continuous ranked probability score (CRPS) inspired by a Kaggle challenge \cite{gneiting2007strictly}. While the first three metrics assess the accuracy of the extracted volumes, the CRPS evaluates the quality of the models' confidence in their predictions. In practice, each participant was asked to calculate the volumes and the cumulative probability distribution for each of the four cardiac structures. The CRPS was then computed for each structure as follows:
\begin{equation}
\label{eq:CRPS}
CRPS = \frac{1}{600N} \sum_{m=1}^{N} \sum_{n=0}^{599} \left( P(y\leq n) - \mathds{1}_{\{n \geq V_m\}}\right)^2,
\end{equation}

where $P$ is the predicted cumulative distribution, $n$ is a value ranging from 0 to 599 (we assume that the estimated volumes do not exceed 599 mL, which is consistent with human physiology, even in the presence of pathologies), $N$ denotes the number of patients in the dataset under evaluation, $V_m$ is the reference volume (in mL) for patient $m$ and $\mathds{1}_{\{n \geq V_m\}}$ is the indicator function, which equals 1 if $n \geq V_m$, and $0$ otherwise. Thus, if a participant was able to estimate a volume along with its confidence value for a given structure, it is straightforward to compute the associated cumulative distribution under the Gaussian assumption, where the volume and confidence values are treated as the mean and standard deviation of the distribution. However, participants may provide distributions that do not follow a Gaussian pattern, based on their own strategy. If a participant did not provide any CRPS value, the cumulative distributions associated with each estimated volume were automatically calculated and corresponded to Heaviside step functions centered on the volumes computed automatically from the segmentation estimates.

Finally, if a participant's MVO prediction was missing for a given patient, no value was assigned to the corresponding geometric and clinical metrics. Consequently, these cases were excluded from the average values reported in the results section table. However, to ensure fairness, we also provide the number of missed cases for each method to assess their sensitivity.

\begin{table*}[t]
\centering
\caption{Summary of the main characteristics of the submitted techniques, including backbone architectures, data augmentation strategies, and confidence estimation approaches. Abbreviations: Rotations (R), Flipping (F), Scaling (S), Deformations (D), Gaussian Noise (GN), Brightness (B), Gamma correction (G). \modif{Methods shown in blue correspond to publicly available implementations and provide direct hyperlinks to the associated code repositories}.}
\begin{adjustbox}{max width=\textwidth}
\begin{tabular}{c c c|cccc|ccc|c|c}
\toprule
\multirow{3}{*}{Method} & 
\multirow{3}{*}{\shortstack{Backbone \\ architecture}} & 
\multirow{3}{*}{\shortstack{Number of \\ parameters}} & 
\multicolumn{8}{c|}{Data augmentation} & 
\multirow{3}{*}{\shortstack{Confidence \\ strategy}} \\ 
 &  &  & \multicolumn{4}{c|}{Spatial augmentations} & 
\multicolumn{3}{c|}{Intensity augmentations} & 
\multirow{2}{*}{Synthesis} &  \\ 
 &  &  & R & F & S & D & GN & B & G &  &  \\
\midrule
\href{https://github.com/lino202/myosaiq_scripts/tree/main}{P1}  & nnUNet & 16.1 M & \checkmark & \checkmark & \checkmark & & \checkmark & \checkmark & \checkmark & \href{https://github.com/lino202/3Dseg/tree/minimal_modifications_for_myosaiq_challenge}{3D GAN} & \href{https://github.com/lino202/nnUNet/tree/master}{Logit calibration} \\ 
P2  & Cascaded 3D UNet & 2.6 M & \checkmark &  & \checkmark & \checkmark &  & \checkmark &  & No & Ensemble with 5 models \\ 
P3  & MedNeXt & 5.6 M & \checkmark & \checkmark & \checkmark & \checkmark & \checkmark & \checkmark & \checkmark & No & Ensemble with 5 models \\ 
P4  & xLSTM-UNet & 1.4 M & \checkmark & \checkmark & \checkmark & \checkmark & \checkmark & \checkmark & \checkmark & No & No \\ 
P5  & Cascaded UNet++ & 50.3 M & \checkmark &  & \checkmark &  &  &  &  & No & Ensemble with 8 models \\ 
\href{https://github.com/Celia-Gjt/threeStage_nnUnet_MYOSAIQ}{P6}  & Cascaded 2D nnUNet & 7.4 M & \checkmark  & \checkmark & \checkmark &  & \checkmark &  & \checkmark & No & Ensemble with 10 models \\ 
\bottomrule
\end{tabular}
\end{adjustbox}
\label{tab:participating-teams}
\end{table*}

\subsubsection{\modif{Leaderboard}}
\label{sec:leaderboard}

\modif{A leaderboard was computed during the first phase of the challenge. Two separate overall rankings were established based on (i) geometric metrics (Dice, ASSD, HD) and (ii) clinical metrics (CC, MAE, LOA, CRPS). For each ranking, a score was computed by summing the ranks obtained by each method across all metrics and cardiac structures (LV, MYO, MI), where ranks ranged from 1 (best) to N (worst), with N denoting the number of participants. MVO was not included in the ranking computation, as the methods do not segment the same number of patients. A dedicated analysis of MVO segmentation performance was, however, conducted and discussed separately during the FIMH 2023.
}

\subsection{Backbones architectures}
\label{sec:backbones-architectures}

All teams employed UNet-inspired backbone architectures, including nnUNet (P1), 3D and 2D UNet variants (P2 and P6), UNet++ (P5), xLSTM-UNet (P4), and MedNeXt (P3). This diversity resulted in a broad spectrum of model complexities, ranging from 1.4 million parameters (P4) to 50.3 million parameters (P5) per base model. \mbox{xLSTM-UNet} features a UNet-based structure enhanced with a parallel cross-window module to capture long-range dependencies \citep{beck_2024}, along with a cross-attention transformer block to compute novel feature representations using cross-attention mechanisms. Team P3 implemented a model based on the MedNeXt architecture \citep{Roy_2023}. To improve performance, each convolutional block was modified with separable convolutional kernels and inverted bottleneck layers, drawing inspiration from the design principles of ConvNeXt \citep{liu_2022}. These modifications aim to maintain rich semantic information across the network. Team P5 used the \mbox{UNet++} \citep{Zhou_2020} architecture with eight different backbones based on the widely available VGG, ResNet, and DenseNet methods. The eight segmentation predictions were fused for each class using a STAPLE algorithm \citep{Warfield_2004}, and the resulting probability maps were entered into a multi-class continuous min-cut algorithm to generate the final segmentation with five classes dimensions. 

Three teams (P2, P5, P6) employed a cascaded strategy where the first stage was used to detect the LV and the full myocardium. This result was then concatenated with the original image to create the input data given to the second stage which focused on the segmentation and the separation between healthy and damaged myocardial tissue. In the case of D8 dataset, a third stage was employed to separate between MI and MVO. Teams P2 and P5 trained their cascaded models in an end-to-end manner, whereas P6 trained each base UNet separately. Note that for these methods, the total number of parameters corresponds to the values presented in Table \ref{tab:participating-teams}, multiplied by the number of levels involved in the cascade strategy.

\subsection{Data augmentation}
\label{sec:data-augmentation}

All participants in the challenge employed various data augmentation techniques to enhance their models. Two main categories of data augmentations were considered: (1) spatial transformations, which increase the sample size by applying rotations, flips, scaling, or deformations to the original images; and (2) intensity-based techniques, which preserve the spatial configuration of anatomical structures while altering their visual appearance. Team P1 employed a 3D GAN-based data augmentation strategy to train their nnUNet model using a combination of original and synthetic data. In particular, a 3D-UNet generator and a PatchGAN discriminator were employed to train the GAN model. Each sample was preprocessed through resampling, padded cropping, and Z-score normalization. For synthetic data generation, the training labels were deformed using both rigid and non-rigid transformations.

\subsection{Confidence strategy}
\label{sec:confidence-strategy}

Five teams developed strategies to enhance the confidence in their clinical metric estimates. Team P1 calibrated their nnU-Net model output using a temperature scaling procedure \citep{guo_2017} to generate probabilities that better reflect true values. These probabilities were then used to compute a volume-wise probability function and its cumulative density function, which served to calculate the CRPS metrics \ref{eq:CRPS}. Teams P2, P3, P5, and P6 employed different ensemble strategies. Teams P2 and P3 trained five models and averaged their outputs to produce the final class predictions for each image. Similarly, team P5 trained eight models but replaced the averaging step with a merging procedure based on the STAPLE algorithm \citep{Warfield_2004}. Team P6 trained ten models to compute the mean and standard deviation maps for each class and image. These maps were then used to define Gaussian distributions of the estimated volumes, which were employed to calculate the CRPS metrics.

\subsection{Foundation models}
\label{sec:foundation-models}

The application of foundation models for medical image segmentation has seen a rapid expansion in recent years, driven by the advent of the SAM model developed for natural image processing \citep{kirillov_2023}. These methods are semi-automatic and require initialization using prompts such as bounding boxes, points, masks, or text. MedSAM \citep{Ma_2024} is one of the most advanced model for medical image segmentation. It was developed on a large-scale medical image dataset with 1.6 M image-mask pairs, covering 10 imaging modalities. This large-scale dataset allows MedSAM to learn a rich representation of medical images, capturing a broad spectrum of anatomies and lesions across different modalities. The dataset has been created by collating images from publicly available medical image segmentation datasets, which were obtained from various sources across the internet. MedSAM was built upon the SAM model by fine-tuning the bounding box prompting and the decoder, resulting in a 91M parameter model. In the context of LGE-MRI, we recently introduced mcMedSAM, a model designed to segment multiple cardiac structures from a single bounding box centered on the left ventricle \citep{goujat25}. This model extends the original MedSAM framework by incorporating structure-specific tokens and a hierarchical end-to-end architecture specifically tailored for multi-class segmentation. We also demonstrated that achieving competitive performance relative to state-of-the-art models such as \mbox{nnU-Net} requires the bounding box to be positioned with high spatial accuracy (within 3 mm in the x and y axes) \citep{goujat25}. \modif{In this study, we removed the bounding-box prompt entirely to evaluate mcMedSAM in a fully automated setting. We froze the encoder parameters inherited from the MedSAM architecture and fine-tuned the decoder on the MYOSAIQ training dataset without data augmentation, enabling a fair comparison with other automatic segmentation methods.} We refer to this configuration as \mbox{mcMedSAM-auto}. For completeness, and to support the discussion, we also report results for the original MedSAM model, whose decoder was fine-tuned on the training dataset \modif{without data augmentation} to better adapt to LGE images. This semi-automatic version, referred to as \mbox{MedSAM-f-semi}, was initialized—during both training and inference—with bounding boxes generated from the reference masks and randomly perturbed along the horizontal and vertical axes by uniform shifts of up to $\pm$1 mm on each side.

\section{Results}

The quality of the results obtained by each method was compared with the inter-observer variability on the testing dataset (denoted as Inter-obs in Tables \ref{tab:geometric_scores} and \ref{tab:clinical_scores}).

\begin{table*}[t]
\centering
\caption{Mean and standard deviations of the geometrical metrics (Dice, ASSD and HD) for the participants P1-P6 and the foundation models (mcMedSAM and MedSAM-f) on the MYOSAIQ challenge test set. \textbf{Bold} face numbers are the best results for the automatic methods in each column and blue numbers indicate results that are not statistically significantly different from the best result in each column (paired t-test, $>0.05$). TP represents the number of times a MVO structure was present in an image and correctly detected. TN represents the number of times a method correctly identified that no MVO structure is present in an image. \modif{Given that the methods do not segment the same number of patients for MVO, we decided not to provide a ranking for this structure.} Inter-obs corresponds to the inter-expert variability computed from two different experts on the same testing dataset.}
\label{tab:geometric_scores}
\resizebox{\textwidth}{!}{
\begin{NiceTabular}{c ccc ccc ccc ccccc}
\toprule
\Block{2-1}{\makecell{Method}} & \Block{1-3}{\makecell{LV}} & & & \Block{1-3}{\makecell{MYO}} & & & \Block{1-3}{\makecell{MI}} & & & \Block{1-5}{\makecell{MVO}}\\
\cmidrule(lr){2-4} \cmidrule(lr){5-7} \cmidrule(lr){8-10} \cmidrule(lr){11-15} 
    & \makecell{Dice} & \makecell{ASSD \\ (mm)} & \makecell{HD \\ (mm)}  & \makecell{Dice} & \makecell{ASSD \\ (mm)} & \makecell{HD \\ (mm)} & \makecell{Dice} & \makecell{ASSD \\ (mm)} & \makecell{HD \\ (mm)} & \makecell{Dice} & \makecell{ASSD \\ (mm)} & \makecell{HD \\ (mm)} & \makecell{TP} & \makecell{TN}\\
\cmidrule(lr){1-15}

P1 & \makecell{0.928 \\ \scalebox{0.8}{$\pm$0.031}} & \makecell{0.5 \\ \scalebox{0.8}{$\pm$0.2}} & \textcolor{blue}{\makecell{7.2 \\ \scalebox{0.8}{$\pm$2.4}}} & \textcolor{blue}{\makecell{0.840 \\ \scalebox{0.8}{$\pm$0.036}}} & \textbf{\makecell{0.4 \\ \scalebox{0.8}{$\pm$0.2}}} & \textbf{\makecell{9.0 \\ \scalebox{0.8}{$\pm$2.8}}} & \makecell{0.669 \\ \scalebox{0.8}{$\pm$0.177}} & \textbf{\makecell{1.0 \\ \scalebox{0.8}{$\pm$1.7}}} & \makecell{19.8 \\ \scalebox{0.8}{$\pm$14.4}} & \makecell{0.537 \\ \scalebox{0.8}{$\pm$0.190}} & \makecell{1.2 \\ \scalebox{0.8}{$\pm$1.0}} & \makecell{19.5 \\ \scalebox{0.8}{$\pm$14.6}} & 10/10 & 15/15 \\

\FineRule{2-15}

P2 & \textbf{\makecell{0.933 \\ \scalebox{0.8}{$\pm$0.029}}} & \textbf{\makecell{0.4 \\ \scalebox{0.8}{$\pm$0.2}}} & \textcolor{blue}{\makecell{6.8 \\ \scalebox{0.8}{$\pm$2.1}}} & \textcolor{blue}{\makecell{0.840 \\ \scalebox{0.8}{$\pm$0.038}}} & \textbf{\makecell{0.4 \\ \scalebox{0.8}{$\pm$0.1}}} & \textbf{\makecell{9.0 \\ \scalebox{0.8}{$\pm$3.0}}} & \textbf{\makecell{0.700 \\ \scalebox{0.8}{$\pm$0.145}}} & \textbf{\makecell{1.0 \\ \scalebox{0.8}{$\pm$1.7}}} & \textbf{\makecell{16.8 \\ \scalebox{0.8}{$\pm$13.2}}} & \makecell{0.681 \\ \scalebox{0.8}{$\pm$0.119}} & \makecell{0.7 \\ \scalebox{0.8}{$\pm$0.4}} & \makecell{14.4 \\ \scalebox{0.8}{$\pm$4.7}} & 9/10 & 15/15 \\

\FineRule{2-15}

P3 & \textcolor{blue}{\makecell{0.930 \\ \scalebox{0.8}{$\pm$0.034}}} & \textcolor{blue}{\makecell{0.5 \\ \scalebox{0.8}{$\pm$0.4}}} & \textcolor{blue}{\makecell{7.7 \\ \scalebox{0.8}{$\pm$5.8}}} & \textbf{\makecell{0.841 \\ \scalebox{0.8}{$\pm$0.041}}} & \textbf{\makecell{0.4 \\ \scalebox{0.8}{$\pm$0.2}}} & \textcolor{blue}{\makecell{9.9 \\ \scalebox{0.8}{$\pm$6.1}}} & \makecell{0.688 \\ \scalebox{0.8}{$\pm$0.168}} & \textbf{\makecell{1.0 \\ \scalebox{0.8}{$\pm$1.3}}} & \textcolor{blue}{\makecell{17.3 \\ \scalebox{0.8}{$\pm$12.4}}} & \makecell{0.648 \\ \scalebox{0.8}{$\pm$0.140}} & \makecell{0.8 \\ \scalebox{0.8}{$\pm$0.4}} & \makecell{14.1 \\ \scalebox{0.8}{$\pm$5.0}} & 8/10 & 15/15 \\

\FineRule{2-15}

P4 & \makecell{0.926 \\ \scalebox{0.8}{$\pm$0.031}} & \makecell{0.5 \\ \scalebox{0.8}{$\pm$0.2}} & \textcolor{blue}{\makecell{7.3 \\ \scalebox{0.8}{$\pm$2.4}}} & \makecell{0.827 \\ \scalebox{0.8}{$\pm$0.043}} & \textbf{\makecell{0.4 \\ \scalebox{0.8}{$\pm$0.1}}} & \textcolor{blue}{\makecell{9.5 \\ \scalebox{0.8}{$\pm$3.3}}} & \makecell{0.660 \\ \scalebox{0.8}{$\pm$0.159}} & \textbf{\makecell{1.0 \\ \scalebox{0.8}{$\pm$1.5}}} & \makecell{19.5 \\ \scalebox{0.8}{$\pm$13.5}} & \makecell{0.647 \\ \scalebox{0.8}{$\pm$0.126}} & \makecell{0.7 \\ \scalebox{0.8}{$\pm$0.3}} & \makecell{13.5 \\ \scalebox{0.8}{$\pm$5.0}} & 8/10 & 15/15 \\

\FineRule{2-15}

P5 & \makecell{0.914 \\ \scalebox{0.8}{$\pm$0.034}} & \makecell{0.6 \\ \scalebox{0.8}{$\pm$0.3}} & \makecell{7.9 \\ \scalebox{0.8}{$\pm$3.2}} & \makecell{0.806 \\ \scalebox{0.8}{$\pm$0.043}} & \textbf{\makecell{0.4 \\ \scalebox{0.8}{$\pm$0.2}}} & \makecell{10.6 \\ \scalebox{0.8}{$\pm$3.4}} & \makecell{0.628 \\ \scalebox{0.8}{$\pm$0.170}} & \textcolor{blue}{\makecell{1.2 \\ \scalebox{0.8}{$\pm$1.8}}} & \makecell{23.9 \\ \scalebox{0.8}{$\pm$14.9}} & \makecell{0.552 \\ \scalebox{0.8}{$\pm$0.186}} & \makecell{1.0 \\ \scalebox{0.8}{$\pm$0.5}} & \makecell{16.2 \\ \scalebox{0.8}{$\pm$3.4}} & 9/10 & 15/15 \\

\FineRule{2-15}

P6 & \textcolor{blue}{\makecell{0.931 \\ \scalebox{0.8}{$\pm$0.030}}} & \textbf{\makecell{0.4 \\ \scalebox{0.8}{$\pm$0.2}}} & \textbf{\makecell{6.7 \\ \scalebox{0.8}{$\pm$2.2}}} & \textcolor{blue}{\makecell{0.839 \\ \scalebox{0.8}{$\pm$0.038}}} & \textbf{\makecell{0.4 \\ \scalebox{0.8}{$\pm$0.1}}} & \textcolor{blue}{\makecell{9.6 \\ \scalebox{0.8}{$\pm$3.8}}} & \makecell{0.661 \\ \scalebox{0.8}{$\pm$0.150}} & \textbf{\makecell{1.0 \\ \scalebox{0.8}{$\pm$1.3}}} & \textcolor{blue}{\makecell{19.7 \\ \scalebox{0.8}{$\pm$13.2}}} & \makecell{0.568 \\ \scalebox{0.8}{$\pm$0.145}} & \makecell{1.1 \\ \scalebox{0.8}{$\pm$0.6}} & \makecell{19.6 \\ \scalebox{0.8}{$\pm$10.3}} & 7/10 & 15/15 \\

\FineRule{2-15}

 \makecell{mcMedSAM \\ \scalebox{0.8}{auto}} & \makecell{0.927 \\ \scalebox{0.8}{$\pm$0.027}} & \textbf{\makecell{0.4 \\ \scalebox{0.8}{$\pm$0.1}}} & \textcolor{blue}{\makecell{7.7 \\ \scalebox{0.8}{$\pm$1.9}}} & \makecell{0.794 \\ \scalebox{0.8}{$\pm$0.041}} & \textbf{\makecell{0.4 \\ \scalebox{0.8}{$\pm$0.1}}} & \makecell{10.1 \\ \scalebox{0.8}{$\pm$3.4}} & \makecell{0.544 \\ \scalebox{0.8}{$\pm$0.167}} & \textcolor{blue}{\makecell{1.3 \\ \scalebox{0.8}{$\pm$1.2}}} & \makecell{27.8 \\ \scalebox{0.8}{$\pm$12.7}} & \makecell{0.328 \\ \scalebox{0.8}{$\pm$0.133}} & \makecell{1.9 \\ \scalebox{0.8}{$\pm$1.1}} & \makecell{26.8 \\ \scalebox{0.8}{$\pm$13.4}} & 10/10 & 15/15 \\

\cmidrule(lr){1-15}

\makecell{mcMedSAM \\ \scalebox{0.8}{semi}} & \makecell{0.965 \\ \scalebox{0.8}{$\pm$0.007}} & \makecell{0.2 \\ \scalebox{0.8}{$\pm$0.0}} & \makecell{4.8 \\ \scalebox{0.8}{$\pm$1.6}} & \makecell{0.890 \\ \scalebox{0.8}{$\pm$0.028}} & \makecell{0.2 \\ \scalebox{0.8}{$\pm$0.1}} & \makecell{7.4 \\ \scalebox{0.8}{$\pm$3.0}} & \makecell{0.740 \\ \scalebox{0.8}{$\pm$0.079}} & \makecell{0.4 \\ \scalebox{0.8}{$\pm$0.2}} & \makecell{8.4 \\ \scalebox{0.8}{$\pm$5.0}} & \makecell{0.696 \\ \scalebox{0.8}{$\pm$0.088}} & \makecell{0.4 \\ \scalebox{0.8}{$\pm$0.1}} & \makecell{5.2 \\ \scalebox{0.8}{$\pm$1.1}} & -- & -- \\

\cmidrule(lr){1-15}
Inter-obs & \makecell{0.954 \\ \scalebox{0.8}{$\pm$0.041}} & \makecell{0.3 \\ \scalebox{0.8}{$\pm$0.3}} & \makecell{5.7 \\ \scalebox{0.8}{$\pm$4.5}} & \makecell{0.902 \\ \scalebox{0.8}{$\pm$0.083}} & \makecell{0.2 \\ \scalebox{0.8}{$\pm$0.2}} & \makecell{7.5 \\ \scalebox{0.8}{$\pm$5.2}} & \makecell{0.906 \\ \scalebox{0.8}{$\pm$0.100}} & \makecell{0.2 \\ \scalebox{0.8}{$\pm$0.3}} & \makecell{9.2 \\ \scalebox{0.8}{$\pm$10.7}} & \makecell{0.937 \\ \scalebox{0.8}{$\pm$0.088}} & \makecell{0.1 \\ \scalebox{0.8}{$\pm$0.1}} & \makecell{2.0 \\ \scalebox{0.8}{$\pm$2.1}} & -- & -- \\

\bottomrule
\end{NiceTabular}
}
\end{table*}

\begin{table*}[t]
\centering
\caption{\modif{Bootstrap analysis of the segmentation performance on the MYOSAIQ challenge test set. For each participant (P1–P6) and automatic version of the foundation model (mcMedSAM), the Dice, ASSD, and HD metrics are reported as \textit{bootstrap mean}, \textit{standard deviation}, and 95\% \textit{confidence interval} (CI), estimated from 5,000 bootstrap resamples of the test subjects. Confidence intervals are reported as [$CI_{\text{lower}}$, $CI_{\text{upper}}$].}}
\label{tab:geometric_scores_bootstrap}
\resizebox{\textwidth}{!}{
\begin{NiceTabular}{c ccc ccc ccc ccccc}
\toprule
\Block{2-1}{\makecell{Method}} & \Block{1-3}{\makecell{LV}} & & & \Block{1-3}{\makecell{MYO}} & & & \Block{1-3}{\makecell{MI}} & & & \Block{1-3}{\makecell{MVO}}\\
\cmidrule(lr){2-4} \cmidrule(lr){5-7} \cmidrule(lr){8-10} \cmidrule(lr){11-13} 
    & \makecell{Dice} & \makecell{ASSD \\ (mm)} & \makecell{HD \\ (mm)}  & \makecell{Dice} & \makecell{ASSD \\ (mm)} & \makecell{HD \\ (mm)} & \makecell{Dice} & \makecell{ASSD \\ (mm)} & \makecell{HD \\ (mm)} & \makecell{Dice} & \makecell{ASSD \\ (mm)} & \makecell{HD \\ (mm)}\\
\cmidrule(lr){1-13}

P1 & \makecell{0.928; 0.004 \\ \scalebox{0.8}{[0.921, 0.935]}} & \makecell{0.5; 0.0 \\ \scalebox{0.8}{[0.4, 0]}} & \makecell{7.2; 0.3 \\ \scalebox{0.8}{[6.7, 7.7]}} & \makecell{0.840; 0.004 \\ \scalebox{0.8}{[0.832, 0.848 ]}} & \makecell{0.4; 0.0 \\ \scalebox{0.8}{[0.3, 0 ]}} & \makecell{9.0; 0.3 \\ \scalebox{0.8}{[8.4, 9.6]}} & \makecell{0.669; 0.02 \\ \scalebox{0.8}{[0.627, 0.706]}} & \makecell{1.0; 0.2 \\ \scalebox{0.8}{[0.7, 0]}} & \makecell{19.8; 1.6 \\ \scalebox{0.8}{[16.8, 23.2]}} & \makecell{0.538; 0.063 \\ \scalebox{0.8}{[0.404, 0.652]}} & \makecell{1.2; 0.3 \\ \scalebox{0.8}{[0.7, 0]}} & \makecell{19.6; 4.6 \\ \scalebox{0.8}{[12.3, 29.5]}} \\

\FineRule{2-13}

P2 & \makecell{0.933; 0.003 \\ \scalebox{0.8}{[0.926, 0.939]}} & \makecell{0.4; 0.0 \\ \scalebox{0.8}{[0.4, 0]}} & \makecell{6.8; 0.2 \\ \scalebox{0.8}{[6.4, 7.3]}} & \makecell{0.840; 0.004 \\ \scalebox{0.8}{[0.831, 0.848 ]}} & \makecell{0.4; 0.0 \\ \scalebox{0.8}{[0.3, 0 ]}} & \makecell{9.0; 0.3 \\ \scalebox{0.8}{[8.4, 9.7]}} & \makecell{0.700; 0.016 \\ \scalebox{0.8}{[0.668, 0.73]}} & \makecell{1.0; 0.2 \\ \scalebox{0.8}{[0.7, 0]}} & \makecell{16.9; 1.5 \\ \scalebox{0.8}{[14.1, 19.9]}} & \makecell{0.681; 0.039 \\ \scalebox{0.8}{[0.601, 0.756]}} & \makecell{0.7; 0.1 \\ \scalebox{0.8}{[0.5, 0]}} & \makecell{14.4; 1.5 \\ \scalebox{0.8}{[11.6, 17.6]}} \\

\FineRule{2-13}

P3 & \makecell{0.930; 0.004 \\ \scalebox{0.8}{[0.922, 0.937]}} & \makecell{0.5; 0.0 \\ \scalebox{0.8}{[0.4, 0]}} & \makecell{7.7; 0.7 \\ \scalebox{0.8}{[6.7, 9.2]}} & \makecell{0.841; 0.005 \\ \scalebox{0.8}{[0.832, 0.85 ]}} & \makecell{0.4; 0.0 \\ \scalebox{0.8}{[0.3, 0 ]}} & \makecell{9.9; 0.7 \\ \scalebox{0.8}{[8.7, 11.4]}} & \makecell{0.688; 0.019 \\ \scalebox{0.8}{[0.648, 0.723]}} & \makecell{1.0; 0.2 \\ \scalebox{0.8}{[0.7, 0]}} & \makecell{17.3; 1.4 \\ \scalebox{0.8}{[14.7, 20.2]}} & \makecell{0.647; 0.05 \\ \scalebox{0.8}{[0.545, 0.741]}} & \makecell{0.8; 0.2 \\ \scalebox{0.8}{[0.5, 0]}} & \makecell{14.1; 1.8 \\ \scalebox{0.8}{[10.9, 17.8]}} \\

\FineRule{2-13}

P4 & \makecell{0.926; 0.003 \\ \scalebox{0.8}{[0.919, 0.933]}} & \makecell{0.5; 0.0 \\ \scalebox{0.8}{[0.4, 0]}} & \makecell{7.3; 0.3 \\ \scalebox{0.8}{[6.8, 7.8]}} & \makecell{0.827; 0.005 \\ \scalebox{0.8}{[0.818, 0.836 ]}} & \makecell{0.4; 0.0 \\ \scalebox{0.8}{[0.4, 0 ]}} & \makecell{9.5; 0.4 \\ \scalebox{0.8}{[8.8, 10.3]}} & \makecell{0.659; 0.018 \\ \scalebox{0.8}{[0.623, 0.693]}} & \makecell{1.0; 0.2 \\ \scalebox{0.8}{[0.7, 0]}} & \makecell{19.5; 1.5 \\ \scalebox{0.8}{[16.6, 22.7]}} & \makecell{0.647; 0.045 \\ \scalebox{0.8}{[0.555, 0.73]}} & \makecell{0.7; 0.1 \\ \scalebox{0.8}{[0.5, 0]}} & \makecell{13.5; 1.8 \\ \scalebox{0.8}{[9.9, 17.0]}} \\

\FineRule{2-13}

P5 & \makecell{0.914; 0.004 \\ \scalebox{0.8}{[0.906, 0.921]}} & \makecell{0.6; 0.0 \\ \scalebox{0.8}{[0.5, 0]}} & \makecell{7.9; 0.4 \\ \scalebox{0.8}{[7.3, 8.6]}} & \makecell{0.806; 0.005 \\ \scalebox{0.8}{[0.797, 0.816 ]}} & \makecell{0.4; 0.0 \\ \scalebox{0.8}{[0.4, 0 ]}} & \makecell{10.6; 0.4 \\ \scalebox{0.8}{[9.9, 11.4]}} & \makecell{0.627; 0.019 \\ \scalebox{0.8}{[0.59, 0.663]}} & \makecell{1.2; 0.2 \\ \scalebox{0.8}{[0.9, 0]}} & \makecell{23.9; 1.7 \\ \scalebox{0.8}{[20.7, 27.2]}} & \makecell{0.553; 0.062 \\ \scalebox{0.8}{[0.431, 0.674]}} & \makecell{1.0; 0.2 \\ \scalebox{0.8}{[0.7, 0]}} & \makecell{16.2; 1.1 \\ \scalebox{0.8}{[14.0, 18.3]}} \\

\FineRule{2-13}

P6 & \makecell{0.931; 0.003 \\ \scalebox{0.8}{[0.924, 0.938]}} & \makecell{0.4; 0.0 \\ \scalebox{0.8}{[0.4, 0]}} & \makecell{6.7; 0.2 \\ \scalebox{0.8}{[6.2, 7.2]}} & \makecell{0.839; 0.004 \\ \scalebox{0.8}{[0.83, 0.847 ]}} & \makecell{0.4; 0.0 \\ \scalebox{0.8}{[0.3, 0 ]}} & \makecell{9.6; 0.4 \\ \scalebox{0.8}{[8.8, 10.4]}} & \makecell{0.661; 0.017 \\ \scalebox{0.8}{[0.626, 0.694]}} & \makecell{1.0; 0.1 \\ \scalebox{0.8}{[0.7, 0]}} & \makecell{19.7; 1.5 \\ \scalebox{0.8}{[16.9, 22.7]}} & \makecell{0.568; 0.054 \\ \scalebox{0.8}{[0.462, 0.67]}} & \makecell{1.1; 0.2 \\ \scalebox{0.8}{[0.7, 0]}} & \makecell{19.7; 3.9 \\ \scalebox{0.8}{[12.9, 27.7]}} \\

\FineRule{2-13}

\makecell{mcMedSAM \\ \scalebox{0.8}{auto}} & \makecell{0.927; 0.003 \\ \scalebox{0.8}{[0.921, 0.933]}} & \makecell{0.4; 0.0 \\ \scalebox{0.8}{[0.4, 0]}} & \makecell{7.7; 0.2 \\ \scalebox{0.8}{[7.2, 8.1]}} & \makecell{0.794; 0.005 \\ \scalebox{0.8}{[0.784, 0.802 ]}} & \makecell{0.4; 0.0 \\ \scalebox{0.8}{[0.4, 0 ]}} & \makecell{10.1; 0.4 \\ \scalebox{0.8}{[9.4, 10.9]}} & \makecell{0.543; 0.019 \\ \scalebox{0.8}{[0.505, 0.579]}} & \makecell{1.3; 0.1 \\ \scalebox{0.8}{[1.1, 0]}} & \makecell{27.8; 1.4 \\ \scalebox{0.8}{[25.1, 30.5]}} & \makecell{0.328; 0.042 \\ \scalebox{0.8}{[0.247, 0.411]}} & \makecell{1.9; 0.3 \\ \scalebox{0.8}{[1.3, 0]}} & \makecell{26.8; 4.2 \\ \scalebox{0.8}{[19.1, 35.3]}}
 \\
\bottomrule
\end{NiceTabular}
}
\end{table*}

\begin{table*}[t]
\centering
\caption{Clinical scores (CC, MAE, LOA, CRPS, see details in section \ref{sec:metrics}) for the participants P1-P6 and the foundation models (mcMedSAM and MedSAM-f) on the MYOSAIQ challenge test set. \textbf{Bold} face numbers indicate the best results for the automatic methods in each column. \modif{Given that the methods do not segment the same number of patients for MVO, we decided not to provide a ranking for this structure.} Inter-obs corresponds to the inter-expert variability computed from two different experts on the same testing dataset.}
\label{tab:clinical_scores}
\resizebox{\textwidth}{!}{
\begin{NiceTabular}{c cccc cccc cccc cccc}
\toprule
\Block{2-1}{\makecell{Method}} & \Block{1-4}{\makecell{LV}} & & & & \Block{1-4}{\makecell{MYO}} & & & & \Block{1-4}{\makecell{MI}} & & & & \Block{1-4}{\makecell{MVO}}\\
\cmidrule(lr){2-5} \cmidrule(lr){6-9} \cmidrule(lr){10-13} \cmidrule(lr){14-17}
    & \makecell{CC} & \makecell{MAE \\ (mL)} & \makecell{LOA \\ (mL)} & \makecell{CRPS} & \makecell{CC} & \makecell{MAE \\ (mL)} & \makecell{LOA \\ (mL)} & \makecell{CRPS} & \makecell{CC} & \makecell{MAE \\ (mL)} & \makecell{LOA \\ (mL)} & \makecell{CRPS} & \makecell{CC} & \makecell{MAE \\ (mL)} & \makecell{LOA \\ (mL)} & \makecell{CRPS} \\
\cmidrule(lr){1-17}
P1 & 0.969 & 8.2 & 23.8 & \textbf{0.002} & \textbf{0.967} & \textbf{8.6} & \textbf{24.5} & \textbf{0.004} & 0.882 & 3.4 & 12.1 & \textbf{0.001} & 0.972 & 2.7 & 6.5 & 0.000 \\

\FineRule{2-17}

P2 & \textbf{0.980} & \textbf{6.7} & \textbf{18.0} & 0.011 & 0.946 & 11.2 & 31.1 & 0.018 & 0.892 & 3.4 & 12.3 & 0.005 & 0.989 & 1.1 & 3.0 & 0.003 \\

\FineRule{2-17}

P3 & 0.978 & 7.4 & 19.6 & 0.012 & 0.924 & 13.1 & 38.1 & 0.022 & 0.888 & 3.2 & 11.7 & 0.006 & 0.990 & 2.6 & 4.9 & 0.006 \\

\FineRule{2-17}

P4 & 0.977 & 7.5 & 20.4 & 0.012 & 0.918 & 12.3 & 38.2 & 0.021 & 0.868 & 3.3 & 12.6 & 0.006 & 0.997 & 1.4 & 2.9 & 0.004 \\

\FineRule{2-17}

P5 & 0.976 & 8.1 & 19.9 & 0.014 & 0.923 & 17.5 & 52.5 & 0.030 & 0.902 & 3.5 & 11.6 & 0.007 & 0.970 & 2.5 & 5.6 & 0.006 \\

\FineRule{2-17}

P6 & \textbf{0.980} & 6.8 & 18.7 & 0.011 & 0.945 & 12.1 & 33.9 & 0.020 & \textbf{0.909} & \textbf{3.1} & \textbf{10.0} & 0.004 & 0.919 & 4.2 & 10.2 & 0.004 \\

\FineRule{2-17}

\makecell{mcMedSAM \\ \scalebox{0.8}{auto}} & 0.977 & 7.3 & 19.1 & 0.012 & 0.927 & 12.8 & 38.6 & 0.022 & 0.821 & 4.2 & 15.3 & 0.008 & 0.569 & 3.7 & 9.2 & 0.007 \\

\cmidrule(lr){1-17}
\makecell{MedSAM-f \\ \scalebox{0.8}{semi}} & 0.999 & 2.4 & 6 & 0.003 & 0.993 & 3.7 & 12.2 & 0.006 & 0.96 & 2.6 & 8.7 & 0.005 & 0.984 & 2 & 6.9 & 0.003 \\

\cmidrule(lr){1-17}
Inter-obs & 0.980 & 6.1 & 18.2 & 0.010 & 0.969 & 10.0 & 27.0 & 0.018 & 0.970 & 1.5 & 6.4 & 0.004 & 0.963 & 1.0 & 5.0 & 0.002 \\
\bottomrule
\end{NiceTabular}
}
\end{table*}

\subsection{MYOSAIQ challenge results}

\begin{figure}[t]
\centerline{\includegraphics[width=\columnwidth]{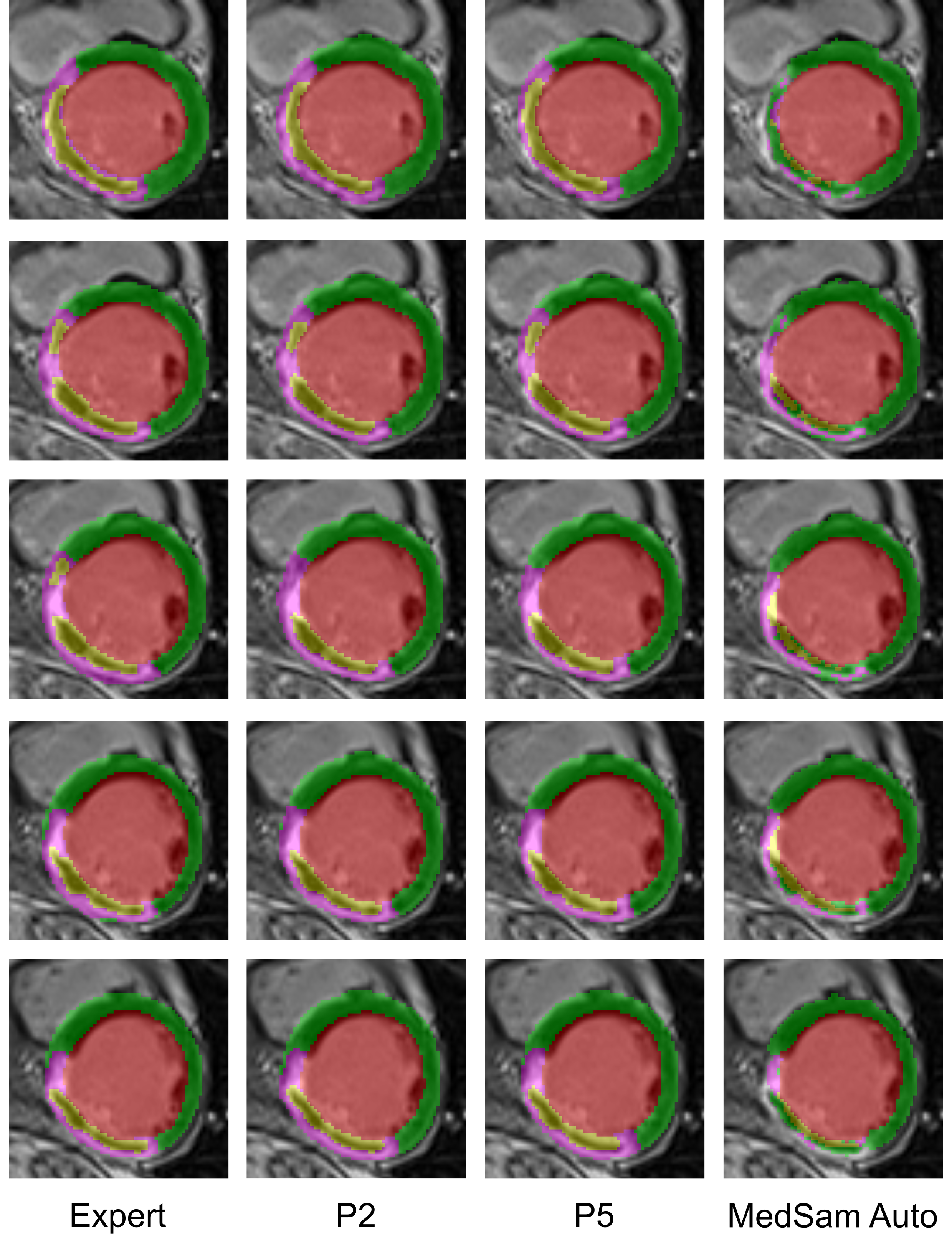}}
 \caption{\modif{Representative example illustrating a typical successful case. P2 and P5 closely match the expert annotation, whereas MedSAM Auto correctly identifies the infarction and microvascular obstruction but with less accurate delineation. Left ventricular cavity (red), healthy myocardium (green), infarction (magenta), and microvascular obstruction (yellow).}}
\label{fig:segmentation-overlapping}
\end{figure}

\begin{figure}[t]
\centerline{\includegraphics[width=\columnwidth]{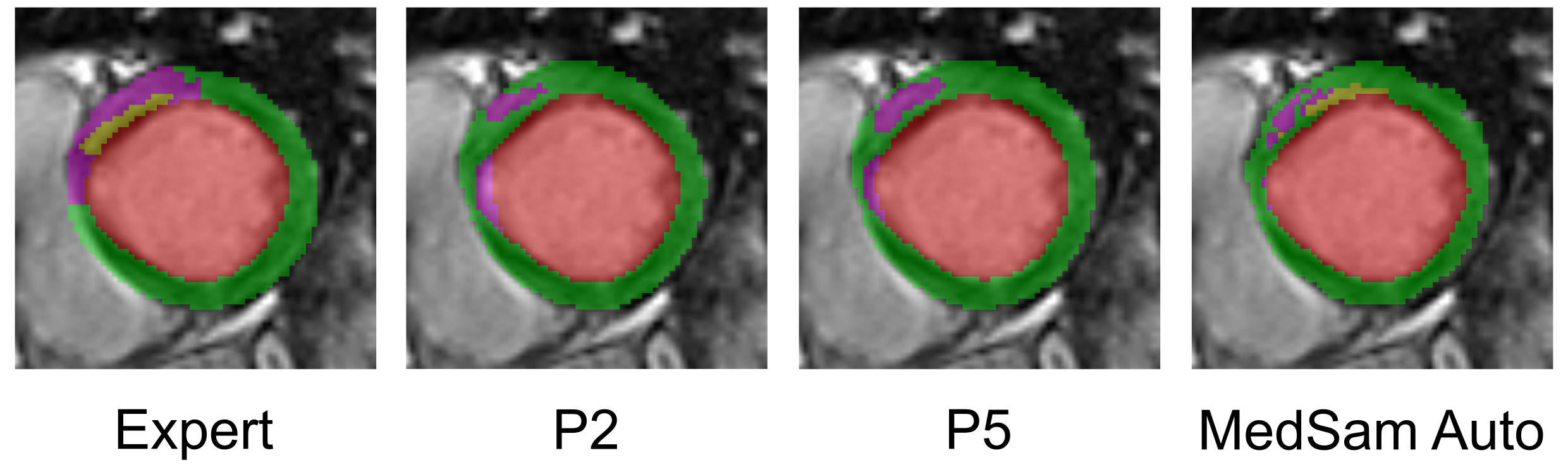}}
 \caption{\modif{Representative challenging case illustrating typical failure modes. P2 and P5 confuse MVO with infarction and miss the expert-annotated infarcted region, whereas MedSAM Auto detects both structures but with localization errors. Left ventricular cavity (red), healthy myocardium (green), infarction (magenta), and microvascular obstruction (yellow).}}
\label{fig:segmentation-overlapping-wrong}
\end{figure}

Table \ref{tab:geometric_scores} and \ref{tab:clinical_scores} display the geometric and clinical scores obtained by all the participants (P1 to P6) and the foundation models (mcMedSAM-auto to mcMedSAM-semi) on the test dataset of the MYOSAIQ challenge. From these tables, we observe that the different participants' methods produced very similar scores, whatever the backbone architecture and the data augmentation strategy. This is further supported by the statistical analysis of the geometrical metrics, which confirms comparable performance levels. Paired t-tests were performed after verifying the normality of the paired differences. \modif{These findings are further supported by the bootstrap analysis reported in Table~\ref{tab:geometric_scores_bootstrap}, where the narrow 95\% confidence intervals indicate that the reported performance estimates are stable with respect to variations in the test cohort.} Interestingly, segmentation performance decreases across the LV, MYO, and MI structures, reflecting increasing task complexity. This is evidenced by the average Dice scores of $0.927$, $0.832$, and $0.668$, respectively, and corresponding HD values of $7.3$, $9.6$, and $19.5$ mm. This trend is also observed in the clinical scores, with average correlation values of $0.977$, $0.937$, and $0.890$. Regarding MVO, the results indicate that these structures are generally detected when present in the image and rarely misidentified when absent, demonstrating a well-balanced performance across methods. However, due to the diffuse nature of MVO, its segmentation remains challenging, as reflected by the average Dice score of $0.606$ and HD score of $16.2$ mm. \modif{Figures \ref{fig:segmentation-overlapping} and \ref{fig:segmentation-overlapping-wrong} illustrate two representative examples of the evaluated methods. The first example corresponds to a typical case where all three methods (P2, P5, and MedSAM Auto) closely match the expert annotation. The second example illustrates a more challenging case, highlighting complementary failure modes. While P2 and P5 tend to confuse microvascular obstruction with infarction and miss the expert-annotated infarcted region, MedSAM Auto successfully differentiates both pathological structures but exhibits less accurate localization.} \modif{It is worth noting that no specific preprocessing tailored to the characteristics of LGE images, nor explicit post-processing steps, were consistently implemented across participants. Therefore, no conclusion can be drawn from this challenge regarding the potential impact or necessity of such pre- or post-processing strategies.}

\begin{figure*}[t]
\centerline{\includegraphics[width=\linewidth]{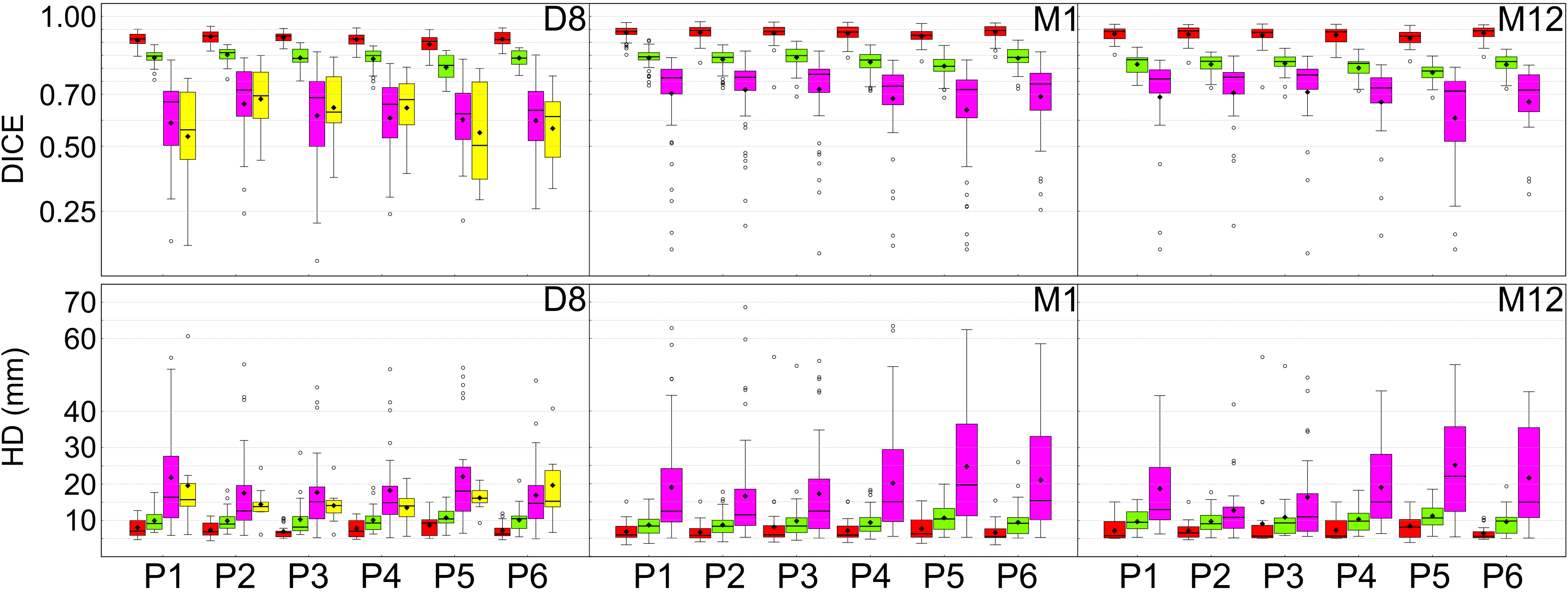}}
\caption{Distribution of Dice and Hausdorff Distance (HD) scores across participants for each label: LV cavity (red), myocardium (green), MI (magenta) and MVO (yellow). The black solid line indicates the median, while the black diamond represents the mean.}
\label{fig:time-effect-on-dice}
\end{figure*}

\subsection{Foundation models' results}

Tables \ref{tab:geometric_scores} and \ref{tab:clinical_scores} report the performance of the two foundation models evaluated in this study. The results show that the fully automatic model achieves competitive performance for the LV, slightly lower performance for the MYO, and clearly inferior performance for the MI and MVO when compared with the CNN-based models. In contrast, the semi-automatic model obtains high scores for the different structures, with geometric accuracy metrics such as a HD score of 4.8 mm for the LV, 7.4 mm for the MYO, 8.4 mm for the MI and 5.2 mm for the MVO, which are the lowest among all the models tested. Although these results are influenced by the initialization procedure, they highlight the critical role of the bounding box in such models and quantify the performance improvements brought by its use.

\subsection{Effect of time elapsed since reperfusion therapy}

Figure \ref{fig:time-effect-on-dice} presents boxplots of Dice and HD scores for each participant based on the time elapsed since reperfusion therapy (day 8, month 1, and month 12). This figure shows that Dice scores remain stable over time for LV and MYO across all participants. A performance drop is observed for the MI region at D8 compared to M1 and M12, both in terms of median values and standard deviations. Similarly, HD scores for LV and MYO remain consistent across participants. However, MI results show greater variability: while teams P1, P2, and P3 maintain stable HD scores over time, teams P4, P5, and P6 exhibit a slight performance decline at M1 and M12. These results indicate the overall stability of the DL methods for the segmentation of LV and MYO structures over time and the increased complexity at D8 of the segmentation of the MI region, likely due to the presence of the MVO structures, which complicate infarct zone detection.

\subsection{Confidence in the clinical scores}

Confidence strategies can be evaluated through the analysis of the CRPS score (see Table \ref{tab:clinical_scores}). The ensemble-based methods (teams P2, P3, P5, and P6) produced similar CRPS scores for each structure, suggesting that the choice of strategy has little impact on confidence estimation in the clinical metrics. Interestingly, these results are comparable to those of team P4, for which no specific confidence estimation strategy was applied in their clinical metrics. This indicates that while ensemble strategies may improve segmentation accuracy, they are not particularly effective in capturing uncertainty. In contrast, team P1 employed a more advanced probabilistic framework based on a temperature-scaling procedure. Their approach proved highly effective, achieving CRPS scores 3 to 7 times lower than those of other participants. This demonstrates both the feasibility and the value of developing methods that are not only accurate but also capable of providing reliable confidence estimates.

\section{Discussion and Conclusion}

\subsection{How far are we from solving the LGE MRI segmentation problem ?}

The results indicate that convolution-based approaches yield comparable geometric scores, which are often not statistically different, regardless of the backbone architecture or data augmentation strategy. Moreover, these methods performed similarly independent on when reperfusion therapy was performed, with some differences for the MI region. \modif{Although these observations require validation on larger cohorts, our results suggest that UNet-inspired backbone architectures may reach a performance plateau when relying solely on image data.} While results obtained on the LV are competitive, it appears that the same level of accuracy is still difficult to obtain for the MYO, MI and MVO structures. It is therefore important to assess the performance of these methods relative to inter-observer variability, which motivated us to obtain manual annotations from a second expert for the 81 LGE images in the testing dataset. These additional annotations allow assessing inter-observer variability which further enhances the comprehensiveness of this study. The inter-observer variability measured on the test (Inter-obs) dataset is reported in the last two rows of Tables \ref{tab:geometric_scores} and \ref{tab:clinical_scores}. Unsurprisingly, it appears that the MYO and MI are regions that are more difficult to annotate than the LV, with a HD score of 5.7, 7.5 and 9.2 mm, respectively. \modif{In terms of segmentation, the comparison between the participants’ scores and inter-observer variability suggests that LV delineation has reached high levels of performance under controlled conditions, although challenges related to variability and generalization across clinical settings remain. However, varying levels of difficulty persist for MYO, MI, and MVO, with MI and MVO being the most challenging due to their lower contrast and higher variability.} Regarding clinical indices, the results show that the best-performing methods are competitive with experts for estimating LV and MYO volumes, although improvements are still needed for MI and MVO. These results clearly highlight the need for further improvements in the quality of segmentation for the MI and MVO regions. In this regard, the computed inter-observer values can serve as benchmarks for the community to strive towards in future developments.

\subsection{Where do foundation models stand ?}

In this study, we compared the methods developed by the challenge participants with state-of-the-art foundation models. The fully automatic mcMedSAM model provided accurate LV segmentations but did not yield improvements for the other structures relative to CNN-based approaches. This outcome is likely due to the limited presence of LGE MR images—and thus the absence of MI, and MVO labels—in the pre-training datasets used for the foundation models. Moreover, the inherent flexibility of transformer-based architectures makes them less effective than CNNs when fine-tuned on relatively small datasets such as MYOSAIQ. In contrast, the high performance achieved by the semi-automatic fine-tuned model suggests that MedSAM’s encoder captures generic and transferable features. This enables efficient adaptation of the decoder (2 million parameters), resulting in performance levels comparable to inter-observer variability across both geometric and clinical metrics. While these findings highlight the potential of foundation models to overcome some of the limitations of convolutional architectures in LGE cardiac segmentation, they also reveal the strong dependence of the semi-automatic model on bounding-box initialization, underscoring a critical challenge for progressing toward competitive fully automatic solutions.

\subsection{Why did temperature scaling outperform ensemble-based uncertainty estimation?}

The temperature-scaling strategy developed by P1 clearly outperformed ensemble-based approaches for uncertainty estimation, as reflected by substantially lower CRPS scores. This superiority can be attributed to both the nature of the uncertainty in this task and the properties of the CRPS metric. Ensemble methods mainly capture epistemic uncertainty arising from model variability; however, in this challenge, most participants trained similar architectures on the same small and relatively homogeneous dataset, leading to limited ensemble diversity and consequently weak uncertainty signals. In contrast, temperature scaling directly improves probability calibration without affecting segmentation accuracy. Since CRPS rewards both calibration and sharpness, improvements in probability calibration result in more reliable confidence estimates and, consequently, lower CRPS values. Furthermore, the task is dominated by aleatoric uncertainty associated with ambiguous tissue boundaries (e.g., MI vs. MVO), for which calibration-based methods appear to be more effective than ensemble disagreement.

\subsection{Can we trust automatic estimation of lesion size post-MI ?}

\begin{figure*}[t]
\centerline{\includegraphics[width=\linewidth]{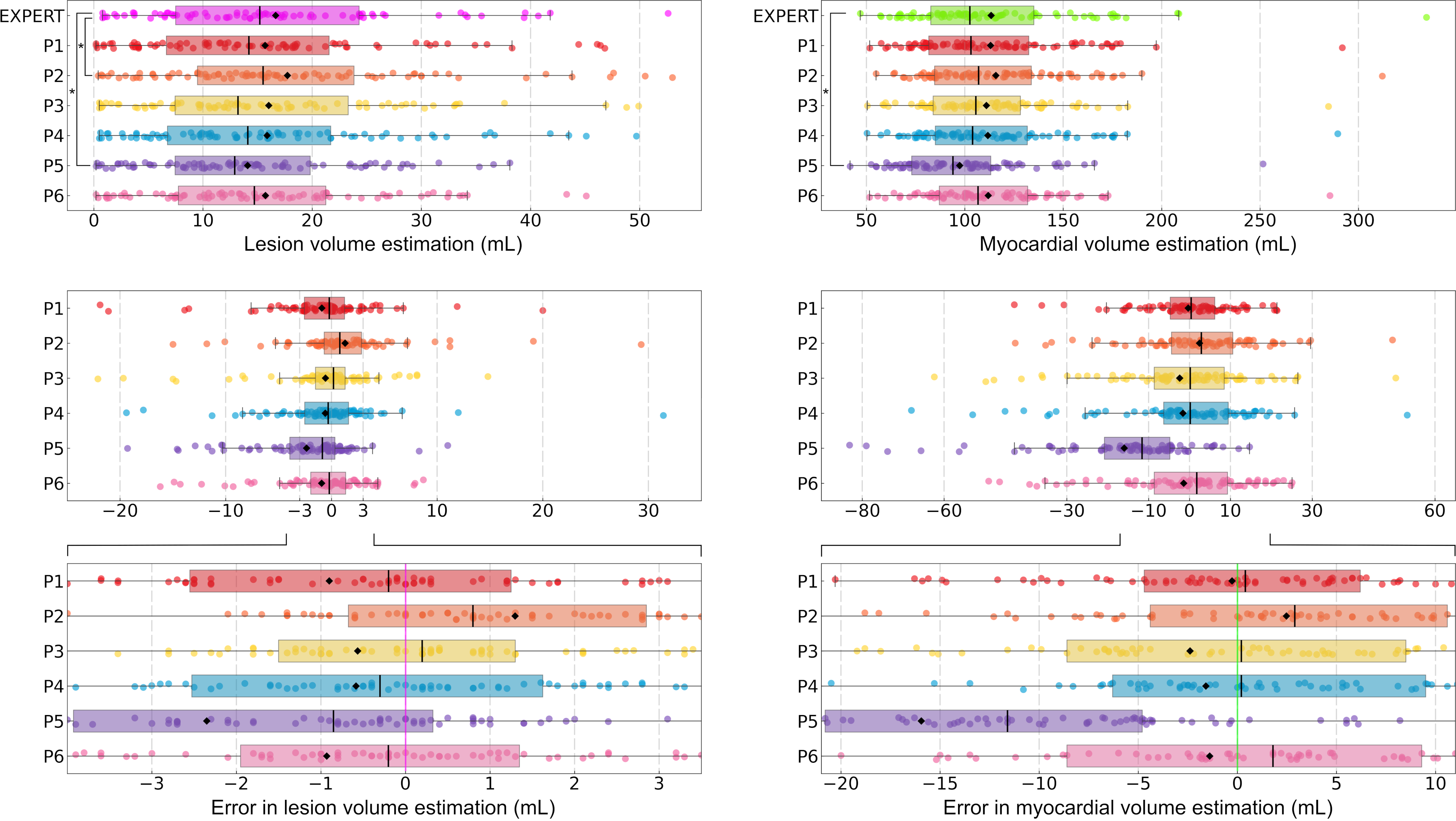}}
\caption{Estimation of lesion (MI) and myocardial (MYO) volumes, along with the distribution of estimation errors across participants. The black line indicates the median, while the black diamond denotes the mean. Paired t-tests were conducted to compare each participant’s volume estimation with the reference derived from expert segmentation. Brackets marked with an asterisk indicate statistically significant differences ($p < 0.05$).
}
\label{fig:lesion-size-distributions}
\end{figure*}

Accurate quantification of myocardial and lesion volumes following myocardial infarction is essential for computing lesion size, expressed as the percentage of infarcted myocardium relative to the total left ventricular myocardial volume. This metric plays a critical role in diagnosis, prognosis, and therapeutic decision-making. Analysis of Tables \ref{tab:geometric_scores} and \ref{tab:clinical_scores} reveals that the methods developed by participants yielded relatively consistent results across both geometric and clinical metrics. Consequently, paired t-tests were performed between the MYO and MI volumes estimated by each participant and the expert reference values. To gain further insight into these results, Figure \ref{fig:lesion-size-distributions} displays the distributions of the estimated volumes, as well as the corresponding errors compared to reference values, for both MYO and MI volumes across participants. Four teams (P1, P3, P4, and P6) produced MYO and MI volume estimates with p-values greater than $0.05$, indicating no statistically significant difference from the expert annotations. Team P2 consistently overestimated the MI volume, despite achieving top geometric scores. Conversely, P5 systematically underestimated both MYO and MI volumes. These systematic deviations likely account for the failure of the statistical tests. These findings underscore the limitations of relying solely on geometric metrics, which capture only part of the clinical relevance. Although additional validation on larger and more heterogeneous cohorts is required, the results indicate that appropriately designed deep learning methods have the potential to produce post-MI lesion size estimates that approach the accuracy of expert assessments.

\subsection{\modif{Generalization across sequence types}}

\modif{To assess the generalization capacity of the different methods, we analyzed their performance according to the acquisition type (2D PSIR vs. 3D-IR-GRE). The dataset consisted of 87\% 3D and 13\% 2D sequences overall, with comparable distributions in the training set (86\% 3D, 14\% 2D) and the test set (93\% 3D, 7\% 2D). Tables \ref{tab:assd_3D} and \ref{tab:assd_2D} report the ASSD scores obtained by P2, P5, and mcMedSAM-auto separately on the 3D-IR-GRE and 2D-PSIR sequences from the test dataset. These results suggest that, despite the imbalance in the training dataset with respect to the sequence type, all methods exhibit good generalization across both acquisition types. The differences in ASSD remain below 0.4 mm for all anatomical structures.}

\begin{table}[t]
\centering
\caption{ASSD scores on the \mbox{3D-IR-GRE} sequences from the test dataset (93\% of the test data)}
\label{tab:assd_3D}
\begin{NiceTabular}{c cccc}
\toprule
Method & LV & MYO & MI & MVO \\
\cmidrule(lr){1-5}

P2 &
\makecell{0.5 \\ \scalebox{0.8}{$\pm$0.2}} &
\makecell{0.4 \\ \scalebox{0.8}{$\pm$0.1}} &
\makecell{1.0 \\ \scalebox{0.8}{$\pm$1.7}} &
\makecell{0.7 \\ \scalebox{0.8}{$\pm$0.3}} \\

\FineRule{2-5}

P5 &
\makecell{0.6 \\ \scalebox{0.8}{$\pm$0.2}} &
\makecell{0.5 \\ \scalebox{0.8}{$\pm$0.2}} &
\makecell{1.2 \\ \scalebox{0.8}{$\pm$1.8}} &
\makecell{1.0 \\ \scalebox{0.8}{$\pm$0.5}} \\

\cmidrule(lr){1-5}

\makecell{mcMedSAM \\ \scalebox{0.8}{auto}} &
\makecell{0.4 \\ \scalebox{0.8}{$\pm$0.1}}  &
\makecell{0.4 \\ \scalebox{0.8}{$\pm$0.1}}  &
\makecell{1.3 \\ \scalebox{0.8}{$\pm$1.2}}  &
\makecell{1.9 \\ \scalebox{0.8}{$\pm$1.1}} \\

\cmidrule(lr){1-5}

Inter-obs &
\makecell{0.3 \\ \scalebox{0.8}{$\pm$0.3}} &
\makecell{0.2 \\ \scalebox{0.8}{$\pm$0.2}} &
\makecell{0.2 \\ \scalebox{0.8}{$\pm$0.3}} &
\makecell{0.1 \\ \scalebox{0.8}{$\pm$0.1}} \\

\bottomrule
\end{NiceTabular}
\end{table}

\begin{table}[t]
\centering
\caption{ASSD scores on the \mbox{2D-PSIR} sequences from the test dataset (7\% of the test data). No MVO cases were present in this subset.}
\label{tab:assd_2D}
\begin{NiceTabular}{c cccc}
\toprule
Method & LV & MYO & MI & MVO \\
\cmidrule(lr){1-5}

P2 &
\makecell{0.3 \\ \scalebox{0.8}{$\pm$0.2}} &
\makecell{0.2 \\ \scalebox{0.8}{$\pm$0.1}} &
\makecell{1.4 \\ \scalebox{0.8}{$\pm$2.1}} & 
\makecell{-- \\ \scalebox{0.8}{--}} \\

\FineRule{2-5}

P5 &
\makecell{0.3 \\ \scalebox{0.8}{$\pm$0.2}} &
\makecell{0.4 \\ \scalebox{0.8}{$\pm$0.1}} &
\makecell{0.8 \\ \scalebox{0.8}{$\pm$0.4}} & 
\makecell{-- \\ \scalebox{0.8}{--}} \\

\cmidrule(lr){1-5}

\makecell{mcMedSAM \\ \scalebox{0.8}{auto}} &
\makecell{0.3 \\ \scalebox{0.8}{$\pm$0.1}}  &
\makecell{0.4 \\ \scalebox{0.8}{$\pm$0.1}}  &
\makecell{2.2 \\ \scalebox{0.8}{$\pm$1.1}}  & 
\makecell{-- \\ \scalebox{0.8}{--}} \\ 

\cmidrule(lr){1-5}

Inter-obs &
\makecell{0.2 \\ \scalebox{0.8}{$\pm$0.2}} & 
\makecell{0.2 \\ \scalebox{0.8}{$\pm$0.1}} &
\makecell{0.1 \\ \scalebox{0.8}{$\pm$0.1}} & 
\makecell{-- \\ \scalebox{0.8}{--}} \\

\bottomrule
\end{NiceTabular}
\end{table}

\subsection{Open science initiative}

In addition to the results and analyses presented in this paper on LGE cardiac MR segmentation, we support open-science principles by making the MYOSAIQ dataset available to the research community under a standard research collaboration agreement. The dataset and manual annotations can be requested through the Human Heart Project website\footnote{https://humanheart-project.creatis.insa-lyon.fr/database/\#collections}. Although the dataset is not directly downloadable due to patient-data governance constraints, access could be granted to academic and industrial researchers upon request motivated by a research project and regulated by a specific consortium agreement. \modif{To further promote reproducibility, two participating teams (P1 and P6), which achieved among the best performances in the challenge, agreed to make their inference code publicly available. Dynamic links to the corresponding repositories are provided in Table~\ref{tab:participating-teams}}. MYOSAIQ is one of the most heterogeneous datasets in LGE cardiac image analysis, encompassing major LGE imaging techniques acquired across 16 centers with three different MR vendors at three distinct post-reperfusion time points. The Codalab platform also remains open for new participants and submissions. We hope that MYOSAIQ will continue to serve as a valuable resource for the community in addressing the remaining challenges highlighted in this study.

\subsection{Limitations}

The definition of a gold standard based on expert annotations remains uncertain. Researchers in this applicative field still debate on which reference methods offer the most reliable baseline in clinical practice, whether $5\sigma$ or $6\sigma$ thresholds compared to remote-myocardium mean values, or approaches such as Full Width at Half Maximum \citep{KARIM201695}. There is also no established methodology for quality assurance or final validation when segmentations come from several experts working on LGE images. Because the absolute accuracy of any segmentation, including those made by experts, cannot be measured, we can only report reproducibility and inter-observer variation limited to two experts in this study. In our work, disagreements between observers were not resolved through consensus or arbitration, particularly at lesion borders where opinions tend to diverge. The inter-observer variability shown here should therefore be viewed as a worst-case estimate.
The results should also be interpreted with caution because they are based on submissions from only six participating teams. The number of samples in the dataset is below 500, which restricts the strength of the conclusions and may not reflect the broader state of the art.
Finally, the MYOSAIQ multi-centre database combines two studies conducted on scanners available at the time. Although it captures a wide range of realistic clinical patterns, it mirrors the scanner market distribution of that period and does not include newer or less common systems. It cannot guarantee fairness across vendors or scanner types, nor does it ensure gender balance. These factors limit how far algorithms trained for this challenge can be generalized. Still, the MYOSAIQ datasets will remain available for future projects that may pool more data and support the development of shared, larger-scale solutions.

\subsection{Conclusions}

This paper presents a comprehensive study on the automatic quantification of myocardial infarction using a multi-center, multi-vendor LGE cardiac MRI dataset. Results were obtained from six teams participating in the MYOSAIQ challenge, as well as from foundation models specifically designed to address the generalization challenge in medical imaging. Inter-observer variability was also assessed to benchmark the performance of each method. \modif{This study confirms that LV segmentation achieves high levels of performance under controlled conditions, while varying levels of difficulty persist for MYO, MI, and MVO, with MI and MVO being the most challenging.} \modif{The comparison across participants suggests that UNet-inspired backbone architectures may reach a performance plateau, with limited benefit from the evaluated data augmentation strategy when relying solely on image data.} Additionally, fine-tuned foundation models appear to have the potential to overcome these limitations, likely due to encoders that have learned generic features, enabling simple yet effective fine-tuning of lightweight decoders. In terms of clinical metrics, deep learning methods achieved competitive correlation, MAE, and LOA scores for LV and MYO volumes comparable to expert variability. Regarding MI lesion size, four out of six teams produced volume estimates that were not significantly different from the expert’s reference values, suggesting the feasibility of automatic measurement of this index in clinical routine.


\acks{This work was supported in part by: the RHU MARVELOUS (ANR-16-RHUS-0009) of l’Université Claude Bernard Lyon 1 (UCBL), within the program "Investissements d'Avenir“ operated by the French National Research Agency (ANR). We also thank Olea Medical and the Virtual Physiological Human Institute (VPHi) for sponsoring the awards for the MYOSAIQ challenge at the FIMH 2023 conference. We also thank the support by AEI (Spain) through projects TED2021-130459B-I00 and PID2022-140556OB-I00 and Aragón Gov. through T39\_23R.}

%
\ethics{The work follows appropriate ethical standards in conducting research and writing the manuscript, following all applicable laws and regulations regarding treatment of animals or human subjects.}

\coi{We declare we don't have conflicts of interest.}

\data{The MYOSAIQ dataset is available to the research community under a standard research collaboration agreement. The dataset and manual annotations can be requested through the Human Heart Project website\footnote{http://humanheart-project.creatis.insa-lyon.fr/}. Although the dataset is not directly downloadable due to patient-data governance constraints, access could be granted to academic and industrial researchers upon request motivated by a research project and regulated by a specific consortium agreement.}

\footnotesize
\bibliography{references}





\end{document}